\documentclass[fleqn,10pt]{wlscirep}
\usepackage[utf8]{inputenc}
\usepackage[T1]{fontenc}

\usepackage{amsmath}
\usepackage{subfigure}
\usepackage{mathptmx} 
 
\DeclareSymbolFont{newfont}{OML}{cmm}{m}{it}
\DeclareMathSymbol{\Epsilon}{3}{newfont}{15}
\DeclareMathSymbol{\Varrho}{3}{newfont}{37}

	\newcommand{\ket}[1]{\left| #1 \right\rangle}

	\newcommand{\braket}[2]{\left\langle #1 \right. \left| #2 \right\rangle}

		\usepackage{color}

\title{Entanglement Protection of Classically Driven Qubits in a Lossy Cavity}

\author[1,2,*]{Alireza Nourmandipour}
\author[2]{Azar Vafafard}
\author[3]{Ali Mortezapour}
\author[2]{Roberto Franzosi}
\affil[1]{Department of Physics, Sirjan University of Technology, 7813733385 Sirjan, Iran}
\affil[2]{QSTAR \& CNR - Istituto Nazionale di Ottica, Largo Enrico Fermi 2, I-50125 Firenze, Italy}
\affil[3]{Department of Physics, University of Guilan, P. O. Box 41335-1914, Rasht, Iran}

\affil[*]{anourmandip@sirjantech.ac.ir}

\begin{abstract}
Quantum technologies able to manipulating single quantum systems, are presently developing. Among the dowries of the quantum realm, entanglement is one of the basic resources for the novel quantum revolution.
Within this context, one is faced with the problem of protecting the entanglement
when a system state is manipulated.
In this paper, we investigate the effect of the classical driving field on the
generation entanglement between two qubits interacting with a bosonic environment. 
We discuss the effect of the classical field on the generation of
entanglement between two (different) qubits and the conditions under which it has a constructive role in protecting the initial-state entanglement from decay induced by its
environment. In particular, in the case of similar qubits, we locate a stationary sub-space
of the system Hilbert space, characterized by states non depending on the environment
properties as well as on the classical driving-field. Thus, we are able to determine the
conditions to achieve maximally entangled stationary states after a transient interaction
with the environment. We show that, overall, the classical driving field has a constructive
role for the entanglement protection in the strong coupling regime. 
Also, we illustrate that a factorable initial-state can be driven in an entangled state and,
even, in an entangled steady-state after the interaction with the environment.
\end{abstract}
\begin{document}

\flushbottom
\maketitle
%
%
\thispagestyle{empty}


\section*{Introduction}

\label{intro}

The rapid experimental progress on quantum control is pushing forward
the second quantum revolution in which quantum technologies able to manipulating
single quantum systems are applied. 
Entanglement is an essential resource in many fields of application for
quantum technologies, for instance in quantum cryptography and computation, in teleportation, in the frequency standard improvement problem, and metrology based on quantum phase estimation \cite{GUHNE20091}. 
In the light of the key role of entanglement, the problem of its protection during the interaction of a system with the surrounding environment represents a basic task to be addressed.
The unavoidable coupling of a quantum system with the surrounding environment brings to the deterioration of the quality of the entanglement or (very often) to its fast destruction.

Despite its key role, entanglement remains elusive and a satisfactory
characterization and quantification of it, is still an open problem \cite{PhysRevA.95.062116, PhysRevA.67.022320}. Different approaches have been developed to classify the entanglement of the variety of states available in the quantum regime \cite{RevModPhys.81.865}.
von Neumann entropy is uniquely accepted as an entanglement measure for pure states of bipartite systems \cite{PhysRevA.56.R3319}.
For mixed states of bipartite systems entanglement of formation \cite{PhysRevLett.80.2245}, entanglement distillation \cite{PhysRevA.54.3824,PhysRevLett.80.5239}, and relative entanglement entropy  \cite{PhysRevLett.78.2275} are largely acknowledged
as faithful measures. In a recent work, \cite{PhysRevA.101.042129}, it has been proposed a measure
of entanglement based on a distance deriving from an adapted application of the Fubini-Study metric, which can be computed for either pure or mixed states of an M-qudit hybrid system.

Besides mathematical proposals for faithful and satisfactory definitions of a multipartite entanglement measure, they have been proposed several schemes for dynamical evolutions of quantum states, able to preserving entanglement from its degrading. These schemes, often, take advantage of phenomena sole preserve of the quantum realm. This is the case, for instance, of schemes with weak measurements \cite{PhysRevA.81.040103,PhysRevA.89.022318} or that use quantum Zeno effect \cite{PhysRevLett.100.090503, PhysRevLett.110.100505, Nourmandipour:16}, quantum error correction \cite{nature432,RevModPhys.87.307}, Stark shift \cite{ABDELATY2007372}. Among various proposed quantum systems for practical implementations, the two-level systems (qubits) have been attracted much great attention duo to their ability of implementation in laboratories \cite{derouault2015exchange,abdel2017some,Zhang_2009,Abdel_Aty_2004}.

On the other hand, the problem of open quantum systems is great of importance due to the presence of dissipation in real systems. This requires new mathematical techniques to model open quantum systems and also to propose new methods to preserve and maybe control the entanglement from deterioration of the surrounding environment. Due to the nonunitary evolution arisen from the interaction of the system with its surrounding environment, modeling open quantum systems seems challenging. For instance, an input-output method \cite{Nourmandipour2015} has been proved to be useful to model such systems. This model is based on the leakage of photons from the non-perfect mirrors of the cavity and leads to a Lorentzian spectral density. This prohibits a definition of a controlling process.

In this work, we propose a scheme where the dynamics cavity-based for two spins, with different transition frequencies, are driven by an external classical field.
The two spins (qubits) interact with the quantized modes of a high-$Q$ cavity and, also, are coupled with an external classical field.
We discuss the effect of the classical field on the generation of entanglement between the two spins, and we explore the conditions under which the classical driving field has a constructive role in protecting the initial-state entanglement from decay induced by its environment. 
In our analysis, the entanglement between the qubits has been quantified with the new multipartite entanglement measure derived in \cite{PhysRevA.101.042129}.
We have compared the entanglement plot achieved with this measure with the one derived by resorting to the concurrence definition, which is appropriate just for bipartite systems \cite{Wootters1998}, obtaining a blatant agreement.

We first consider the case of qubits with the same transition frequency and we show the existence of a stationary sub-space of the Hilbert space of the system. Surprisingly, the stationary states do not depend on the environmental properties as well as on the classical driving field.
Base on this fact, we determine the situation in which a maximally entangled stationary state can be obtained even after interaction with the environment.

Besides, for two similar qubits, we investigate the impact of the classical driving field on the entanglement dynamics. We will illustrate that in the absence of detuning and for any initially entangled state, the entanglement decays in time. However, in the strong coupling regime, we observe oscillatory behavior of the entanglement due to the memory depth of the environment.  We show that the classical driving field, by suppressing such oscillatory behaviors, has a constructive role for entanglement protection in the strong coupling regime.
We also illustrate that an initially factorable state can be entangled and its entanglement can even persist at a steady state via interaction with the common environment.
Also, this stationary value of entanglement can be exceeded in the presence of the classical driving field, for instance, in the strong coupling regime and for sufficiently large values of the amplitude of the classical driving field, a high degree of entanglement can be achieved from an initially disentangled state.
Finally, the role of detuning on the survival of initial entanglement in both weak and strong coupling regimes is investigated and we find that overall the detuning has a constructive role in the preservation of entanglement.

\section*{The Model}
\label{ModSol}

\begin{figure}[ht]
   \centering
\includegraphics[width=0.6\textwidth]{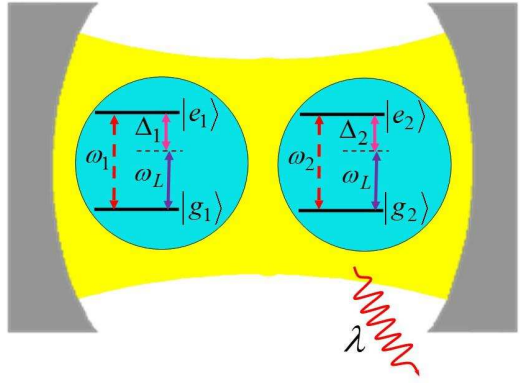}
   \caption{\label{Fig1} Schematic illustration of a setup in which the two-qubit system is driving by a classical field inside a leaky cavity.}
  \end{figure}

We consider a system in which two qubits with (different) transition frequencies $\omega_j$ ($j=1$ and $2$) are driven by an external classical field. We also assume that the qubits are also interacting with a common zero-temperature environment formed by the quantized modes of a high-Q cavity, as illustrated in Fig. \ref{Fig1}. The Hamiltonian describing the whole system in the dipole and rotating wave approximations is written as (we assume $\hbar=1$)
\begin{equation}
\label{Eq1}
\begin{aligned}
\hat{H}=\sum_{j=1}^{2}\frac{\omega_j}{2}\hat{\sigma}_z^{(j)}+\sum_{k}\omega_{k} {\hat{a}_{k}}^{\dagger}\hat{a}_{k}+\sum_{j=1}^{2}\left( \sum_{k}\alpha_jg_{k}\hat{a}_{k}\hat{\sigma}_{+}^{(j)}+\Omega e^{-i\omega _Lt}\hat{\sigma}_{+}^{(j)}+\text{H.c.}\right) \, ,
\end{aligned}
\end{equation}
where ${{\hat{\sigma}}_{z}^{(j)}}=\left| e \right\rangle _j\left\langle  e \right|-\left| g \right\rangle_j \left\langle  g \right|$ is the population inversion operator of the $j$th qubit with transition frequency $\omega_j$, ${{\omega }_{L}}$ and ${{\omega }_{k}}$  represent the frequencies of  the classical driving field and the cavity quantized modes, respectively. ${{\hat{\sigma }}_{+}^{(j)}}=\left| e \right\rangle_j\left\langle  g \right|$ (${{\hat{\sigma }}_{-}^{(j)}}=\left| g \right\rangle_j\left\langle  e \right|$) denotes the raising (lowering) operator for the $j$th qubit, while ${{\hat{a}}_{k}}$ ($\hat{a}_{k}^{\dagger}$) are the annihilation (creation) operators of the cavity $k$th mode. In addition, $\Omega $ and ${{g}_{k}}$ represent the coupling strength of the interactions of the qubits with the classical driving field and the cavity modes, respectively. Furthermore, $\alpha_j$ is a dimensionless parameter which measures the interaction of the $j$th qubit with its surrounding environment. We assume that $\Omega $ is a real number and to be small compared to the atomic and laser frequencies ($\Omega <<{{\omega }_{j}},{{\omega }_{L}}$).

A unitary transformation does not change either the eigenvalues of the system
and the degree of entanglement \cite{FEI200577}. 
Therefore, we consider the unitary transformation
$U=e^{-i\omega_L(\hat{\sigma}_z^{(1)}+\hat{\sigma}_z^{(2)})t/2}$,
under which, the Hamiltonian of the system in the rotating reference frame can be written as
\begin{equation}
\label{Eq2}
\begin{aligned}
\hat{H}_{\text{eff}}&=\hat{H}_{\text{I}}+\hat{H}_{\text{II}} \, , \\
\hat{H}_{\text{I}}&=\left( \frac{\Delta_1}{2}\hat{\sigma}_{z}^{(1)}+\frac{\Delta_2}{2}\hat{\sigma}_{z}^{(2)}\right) +\Omega(\hat{\sigma}_{x}^{(1)}+\hat{\sigma}_{x}^{(2)}) \, , \\
\hat{H}_{\text{II}}&=\sum_{k}\omega_k\hat{a}_k^{\dagger}\hat{a}_k+\sum_{j=1}^{2}\left( \sum_{k}\alpha_jg_{k}\hat{a}_{k}\hat{\sigma}_{+}^{(j)}e^{i\omega _Lt}+\text{H.c.}\right) \, ,
\end{aligned}
\end{equation}
in which $\Delta_j={{\omega }_{j}}-{{\omega }_{L}}$ denotes the detuning between the qubit and the classical driving field. A glance at Eq. \eqref{Eq2} reveals that the Hamiltonian $\hat{H}_{\text{I}}$ can be written as $\hat{H}_{\text{I}}=\hat{H}_{\text{I}}^{(1)}+\hat{H}_{\text{I}}^{(2)}$ in which $\hat{H}_{\text{I}}^{(j)}=\frac{\Delta_j}{2}\hat{\sigma}_{z}^{(j)}+\Omega\hat{\sigma}_{x}^{(j)}$, with  $j=1$ and $2$. Each term can be diagonalized by introducing the dressed bases
\begin{equation}
\ket{E}_j=\sin\frac{\eta_j}{2}\ket{g}_j+\cos\frac{\eta_j}{2}\ket{e}_j \ \ \ \text{and} \ \ \  \ket{G}_j=\cos\frac{\eta_j}{2}\ket{g}_j-\sin\frac{\eta_j}{2}\ket{e}_j,
\end{equation}
which are the eigenstates of ${{\hat{H}}_{\text{I}}}^{(j)}$, since it 
results ${{\hat{H}}_{\text{I}}}^{(j)}=\frac{\chi_j}{2}{{\hat{\Varrho}}_{z}}^{(j)}$ in which ${{\hat{\Varrho }}_{z}}^{(j)}=\left| E \right\rangle_j\left\langle  E \right|-\left| G \right\rangle_j\left\langle  G \right|$ and $\chi_j=\sqrt{{{\Delta}_j^{2}}+4{{\Omega }^{2}}}$. In the above relation, $\eta_j =\operatorname{Arctan}[2\Omega /\Delta_j ]$. In this regard, the effective Hamiltonian can be re-written as
\begin{equation}
\label{Heff}
\begin{aligned}
\hat{H}_{\text{eff}}&=\frac{\chi_1}{2}\hat{\Varrho}_{z}^{(1)}+\frac{\chi_2}{2}\hat{\Varrho}_{z}^{(2)}+\sum_{k}\omega _{k}\hat{a}_{k}^{\dagger}\hat{a}_{k}\\
&+\left( \alpha_1{{\cos }^{2}}(\eta_1 /2)\hat{\Varrho}_{+}^{(1)}+\alpha_2{{\cos }^{2}}(\eta_2 /2)\hat{\Varrho}_{+}^{(2)}\right) e^{+i{{\omega }_{L}}t}\sum_{k}{g}_{k}\hat{a}_{k}+\text{H.c.} \, .
\end{aligned}
\end{equation}
Here  ${{\hat{\Varrho }}_{+}}^{(j)}=\left| E \right\rangle_j\left\langle  G \right|$ (${{\hat{\Varrho }}_{-}}^{(j)}=\left| G \right\rangle_j\left\langle  E \right|$) represents the new lowering (raising) operator. It  should be noted that during the derivation of the effective Hamiltonian \eqref{Heff}, the non-conservation energy terms have been neglected  according to the rotating-wave approximation \cite{mortezapour2020effect,Huang2017}.  It should be noted that the counter-rotating terms arising from the unitary transformation have been neglected.

In what follows, it will be proven that it may be useful introducing the collective coupling constant $\alpha_{_T}=(\alpha_1^2+\alpha_2^2)^{1/2}$ and the relative parameters $r_j=\alpha_j/\alpha_{_T}$. There is no need to prove that $r_1^2+r_2^2=1$, therefore, only one parameter, let us say $r_1$, can be taken as an independent variable. Moreover, the weak and strong coupling regimes are explored by varying $\alpha_{_T}$.

Now we are in a position to examine the effect of the classical field on the entanglement dynamics of the two-qubit system. In the new dressed states, we assume that the initial state of the whole system is
\begin{equation}
\ket{\psi_0}=\left( \cos(\theta/2)\ket{E}\ket{G}+\sin(\theta/2) e^{i\phi}\ket{G}\ket{E}\right)\otimes\ket{\boldsymbol{0}}_{R} \, ,
\label{eq:initialstate}
\end{equation}
in which $\ket{\boldsymbol{0}}_{R}=\hat{a}_k\ket{\boldsymbol{1}_{k^{'}}}\delta_{kk^{'}}$ is the multi-mode vacuum state, where $\ket{\boldsymbol{1}_{k}}$ is the multi-mode state representing one photon at frequency $k$ and vacuum state in all other modes. The initial state \eqref{eq:initialstate} evolves into the following state
\begin{equation}
\begin{aligned}
\ket{\psi(t)}&=C_1(t)\ket{E}\ket{G}\ket{\boldsymbol{0}}_{R}+C_2(t)\ket{G}\ket{E}\ket{\boldsymbol{0}}_{R}+\sum_{k}C_{k}(t)\ket{G}\ket{G}\ket{\boldsymbol{1}_k} \, ,
\end{aligned}
\label{eq:state}
\end{equation}
After tracing over the reservoir degrees of freedom, the quantum state of the two-qubit system in the basis $\ket{E}\ket{E}$, $\ket{E}\ket{G}$, $\ket{G}\ket{E}$ and $\ket{G}\ket{G}$ is obtained as
\begin{equation}
 \hat{\rho}(t)= \left( \begin{array}{cccc}
0 & 0 & 0 & 0 \\
0 & \left| C_1(t)\right| ^2 & C_1(t)C_2^{*}(t) & 0 \\
0 &  C_1^{*}(t)C_2(t) &  \left| C_2(t)\right| ^2 &0 \\
0 & 0 & 0 & 1-\left| C_1(t)\right| ^2-\left| C_2(t)\right|^2 \end{array} \right) \, ,
\label{eq:densitymatrix1}
\end{equation}
Substituting Eq. \eqref{eq:state} into the Schr\"{o}dinger equation $\left( i\dot{\ket{\psi}}=\hat{H}\ket{\psi}\right)$ 
we obtain the following  integro-differential equations for  amplitudes for $C_1(t)$, $C_2(t)$ and $C_{k}(t)$
\begin{subequations}
\begin{eqnarray}
\dot{C}_j(t)&=&-i\alpha_j{{\cos }^{2}}(\eta_j /2)e^{i\omega_Lt}\sum_{k}g_kC_k(t)e^{-i(\omega_k-\chi_j)t} \ \ \text{for} \ j=1, 2 \, , \label{eq:diff1}\\
\dot{C}_{k}(t)&=&-ig_k^{*}e^{-i\omega_Lt}\left( \alpha_1{{\cos }^{2}}(\eta_1 /2)e^{i(\omega_k-\chi_1)t}C_1(t)+\alpha_2{{\cos }^{2}}(\eta_2 /2)e^{i(\omega_k-\chi_2)t}C_2(t) \right) \, . \label{eq:diff2}
\end{eqnarray}
\end{subequations}
By substituting into Eq. (\ref{eq:diff1}) the integration of Eq. (\ref{eq:diff2}), we get the following two integro-differential equations for $C_{1}(t)$ and $C_{2}(t)$

\begin{subequations}
\label{eq:newdotu}
\begin{eqnarray}
&\!\!\!\!\!\!\!\!\!\dot{C}_1(t)=-\int_0^t\text{d}t'G(t-t')\left( \alpha_1^2{{\cos }^{4}}(\eta_1 /2)C_1(t') + \alpha_1\alpha_2{{\cos }^{2}}(\eta_1 /2){{\cos }^{2}}(\eta_2 /2)e^{-i\xi_{21}t'}C_2(t')\right)e^{i\xi_1(t-t')} & \, , \label{eq:newu1} \\
&\!\!\!\!\!\!\!\!\!\dot{C}_2(t)=-\int_0^t\text{d}t'G(t-t')\left(  \alpha_1\alpha_2{{\cos }^{2}}(\eta_1 /2){{\cos }^{2}}(\eta_2 /2)e^{i\xi_{21}t'}C_1(t')+\alpha_2^2{{\cos }^{4}}(\eta_2 /2)C_2(t')\right)e^{i\xi_2(t-t')} &\, , \label{eq:newu2} 
\end{eqnarray}
\end{subequations}
in which the kernel $G(t-t')$, is the correlation function defined in terms of continuous limits of the environment frequency as
\begin{eqnarray}\label{Correlation function}
G(t-t')=\int \text{d}\omega J(\omega)e^{i(\omega_L+\omega_c-\omega)(t-t')} \, ,
\end{eqnarray}
where $J(\omega)=W^2 \lambda/\pi[(\omega-\omega_{c})^2+\lambda^{2}]$ is the Lorentzian spectral density. Here, $W$ is proportional to the vacuum Rabi frequency (${\cal R}=\alpha_TW$) and $\lambda$ describes the cavity losses (the rate at which the photons escape the cavity). In the above relations, $\omega_c$ is the fundamental frequency of the cavity, $\xi_j=\chi_j-\omega_c$ and $\xi_{21}=\chi_2-\chi_1$. 

In principle, the above set of coupled differential equations can be solved analytically using, for instance, the Laplace transformation method. However, as will be clarified, only for the special case $\omega_1=\omega_2$ a simple analytical solution can be derived for the amplitude coefficients.

\section*{Similar qubits}
In this section, we consider a special case in which two qubits have the same transition frequency, i.e., $\omega_1=\omega_2\equiv\omega_0$, therefore, $\chi_1=\chi_2\equiv\chi$, $\Delta_1=\Delta_2\equiv\Delta$, $\eta_1=\eta_2\equiv\eta$ and $\xi_1=\xi_2\equiv\xi$. In this situation, the set of equations (\ref{eq:newdotu}) reduces to
\begin{equation}
\label{eq:dotu}
\dot{C}_i(t)=-{{\cos }^{4}}(\eta /2)\int_{0}^{t}\text{d}t'F(t-t')\left( \alpha_i^2C_i(t')+\alpha_i\alpha_j C_j(t')\right), \ \ \ j\neq i 
\end{equation}
in which $F(t-t')=G(t-t')e^{i\xi(t-t')}\, .$

Before considering the dynamical behaviour of the entanglement, it is interesting to first search for a stationary solution of the set of equations (\ref{eq:dotu}), i.e., $C_j(t\rightarrow\infty)$.  This can be done by setting $\dot{C}_j=0$ in (\ref{eq:dotu}) which leads to the following long-living decoherence-free (or sub-radiant) state (after normalization)
\begin{equation}
\ket{\psi_-}=r_2\ket{E}\ket{G}-r_1\ket{G}\ket{E} \, .
\label{eq:subradstate}
\end{equation}
As it is seen, the existence of the subradiant state does not depend on the classical driving field, the form of the spectral density and neither on the resonance/off-resonance condition. It is clear that for $\omega_1\neq\omega_2$ there is no decoherence-free state. It is also worth mentioning that if qubits were illuminated with different classical laser fields, there would not exist a sub-radiant state. The initial state of the system of qubits consists of two parts: the sub-radiation state and its orthogonal state, i.e., super-radiant state $\ket{\psi_+}$, which on the contrary to the sub-radiant state, evolves in time with the survival amplitude ${\cal G}(t)$ with the following equation of motion
  \begin{equation}
  \label{intgdiff}
 \dot{{\cal G}}(t)= -\alpha_{_T}^2{{\cos }^{4}}(\eta /2)\int_{0}^{t}\! F(t-t'){\cal G}(t{'})  \, \mathrm{d}t' \, .
  \end{equation}
Let us assume for a while that the analytical expression for ${{\cal G}}(t)$ is obtained. Then, it is straightforward to show that the amplitudes $C_i(t)$ result \cite{Nourmandipour2015}
\begin{equation}
\label{eq:C}
C_{1}(t)=r_2\beta_-+r_1{\cal G}(t)\beta_+ \ \ \text{and} \ \ C_{2}(t)=-r_1\beta_-+r_2{\cal G}(t)\beta_+.
\end{equation}
where we have introduced the definition $\beta_{\pm}\equiv\braket{\psi_{\pm}}{\psi(0)}$.

\subsection*{Stationary States}

It should be noticed that the amplitude ${\cal G}(t)$ tends to zero at sufficiently long times. This can be realized because it represents the part of the wave function which decays in time. More specifically, according to Eq. (\ref{intgdiff}), at sufficiently long times, we have $\dot{{\cal G}}(t)=0$, which in turns implies ${\cal G}(t\rightarrow\infty)\mapsto 0$. In this situation, the amplitudes (\ref{eq:C}) become
\begin{equation}
\label{eq:Cst}
C_{1}(\infty)=r_2\beta_-, \ \ \ \ \ \   C_{2}(\infty)=-r_1\beta_- .
\end{equation}
According to the above relations and the density matrix (\ref{eq:densitymatrix1}), the stationary solution depends only on the parameters $\beta_-,r_1$ and $r_2$. It does not depend on the classical driving field as well as the environmental parameters. This can be explained as follows: First of all, the two qubits are dissipating into a global (common) environment.  This global dissipation searches for the excitations of the qubits (one or two) in the symmetric subspace, which is then followed by a decay with a fixed rate. The global dissipation leaves the state $\ket{\psi_-}$ fixed.
The decay process can be represented by a Markov link  $\ket{E}\ket{E}\rightarrow\ket{\psi_+}\rightarrow \ket{G}\ket{G}$. It is worth noticing that our model is much more general than the case in which two qubits are symmetrically coupled to the global environment without any driving field (i.e., $r_1=r_2$) \cite{Nourmandipour2016QIC}. Also, we have shown that the stationary state does not depend on the Markovianity of the process.

In order to quantify the degree of entanglement between the two qubits, we use the entanglement measure proposed by one of us very recently in \cite{PhysRevA.101.042129}. It has the advantage of being able to quantify the entanglement of a $M$-qudit hybrid system. It is defined as
\begin{equation}
E(\rho(t)):=\sum_{\mu=0}^{M-1}\left[ \frac{2(d_{\mu}-1)}{d_{\mu}}-\sum_{k=1}^{d_{\mu}^2-1}\text{Tr}\left( \sigma_k^{\mu}\rho(t)\right)^2 \right] ,
\label{eq:entmea}
\end{equation}
in which $M$ is the number of the subsystems and $d_{\mu}$ is the dimension of the Hilbert space of $\mu$th subsystem. Here $M=2$ and $d_0=d_1=2$. Usually it is convenient to normalize this measure with respect to the number of subsystems, i.e., $E(\rho(t))/M$. In this regard, $0\leq E(\rho(t))/M\leq 1$. The more value of this parameter, the more amount of entanglement. 

{ For steady state, the entanglement measure (\ref{eq:entmea}) becomes 

\begin{equation}
E(\rho(\infty))/M=4(r_1r_2)^2|\beta_-|^4.
\label{eq:Const}
\end{equation}
}
A first glance at the above relation reveals an interesting result. It is straightforward to observe that the stationary entanglement is always at its maximum value for symmetric coupling i.e., $r_1=r_2=\frac{1}{\sqrt{2}}$ and $|\beta_-|^2=1$. This situation corresponds to the case that the initial state for the qubits is $\ket{\psi(0)}=\ket{\psi_-}=\frac{1}{\sqrt{2}}\left( \ket{E}\ket{G}-\ket{G}\ket{E}\right)$, in this case, there is only the sub-radiant state into the initial state, which does not decay in time. Therefore, a maximally entangled state as the stationary state is expected. 
\begin{figure}[ht]
\centering
\subfigure{\label{Fig2a}} {\includegraphics[width=0.45\textwidth]{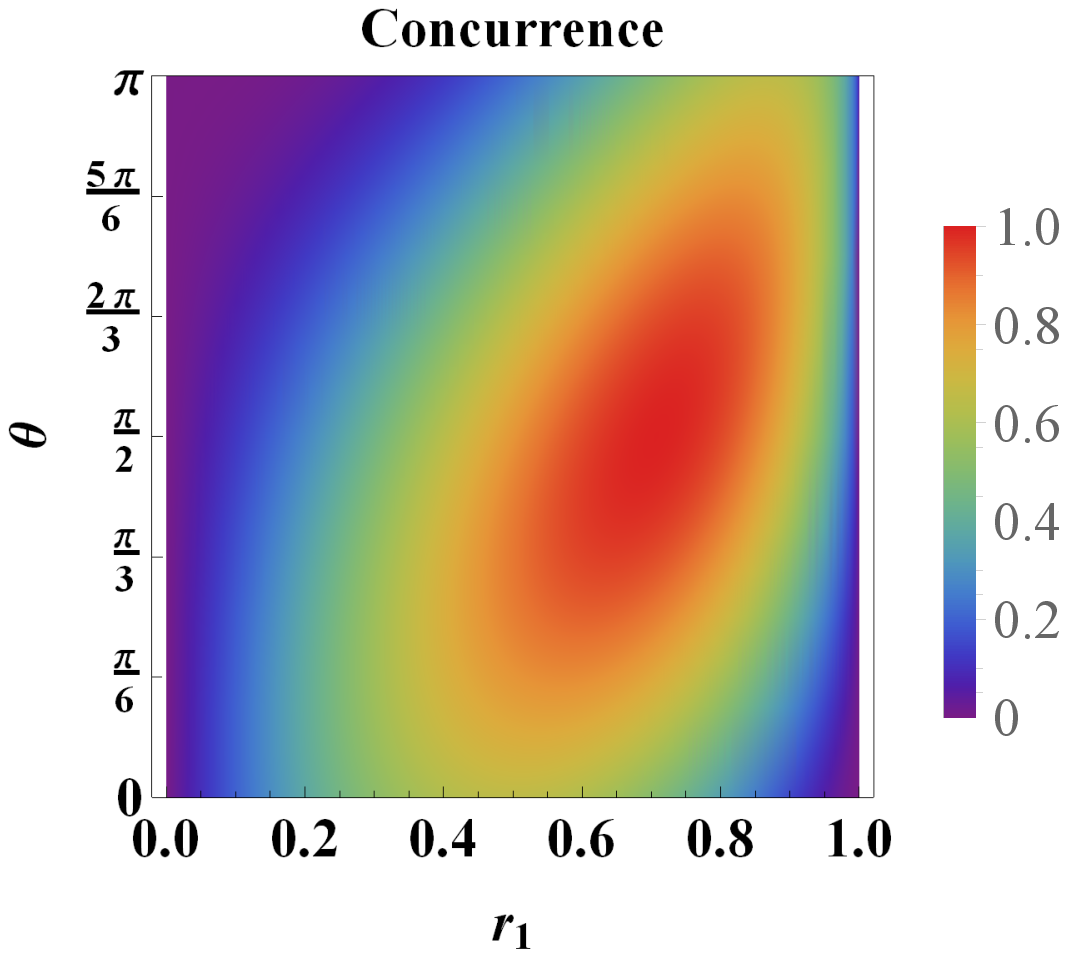}}
\hspace{0.01\textwidth}
\subfigure{\label{Fig2b}} {\includegraphics[width=0.45\textwidth]{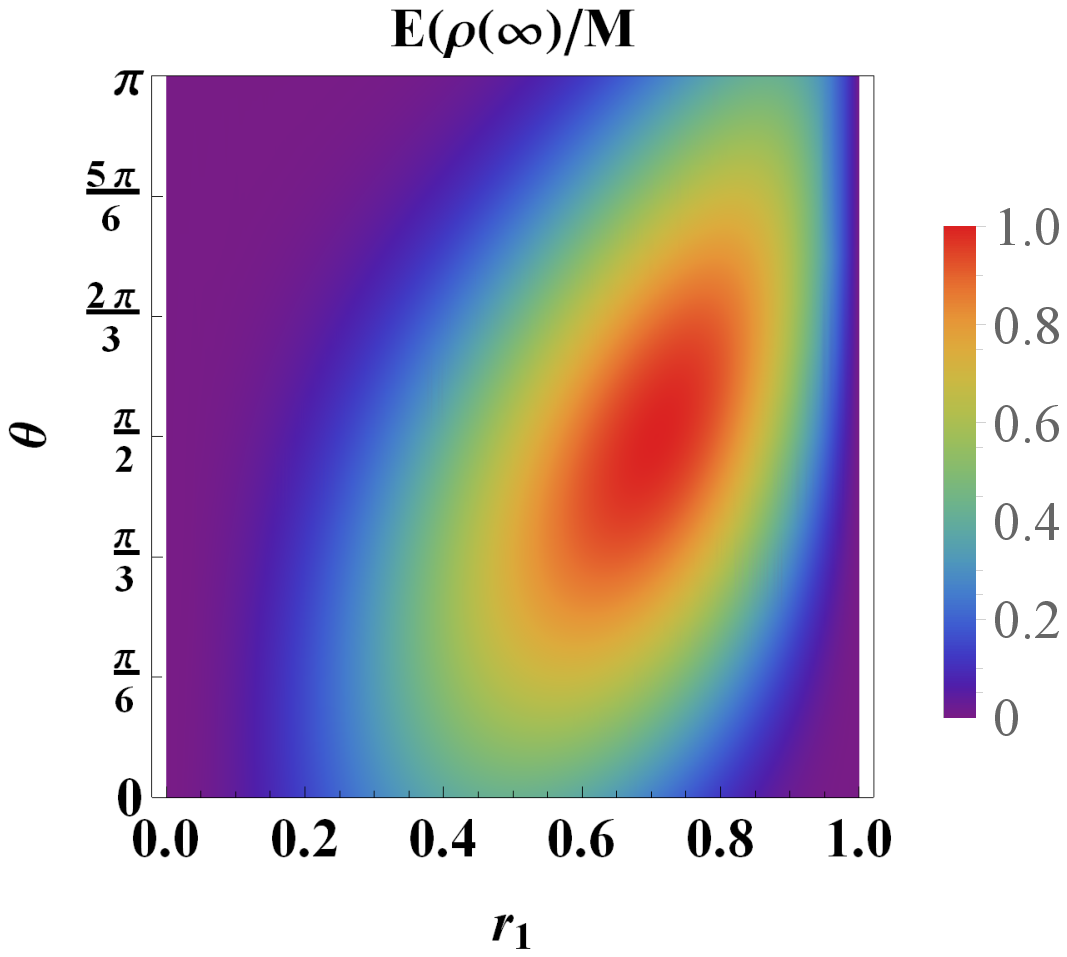}}
\hspace{0.01\textwidth}
\subfigure{\label{Fig2c}} {\includegraphics[width=0.45\textwidth]{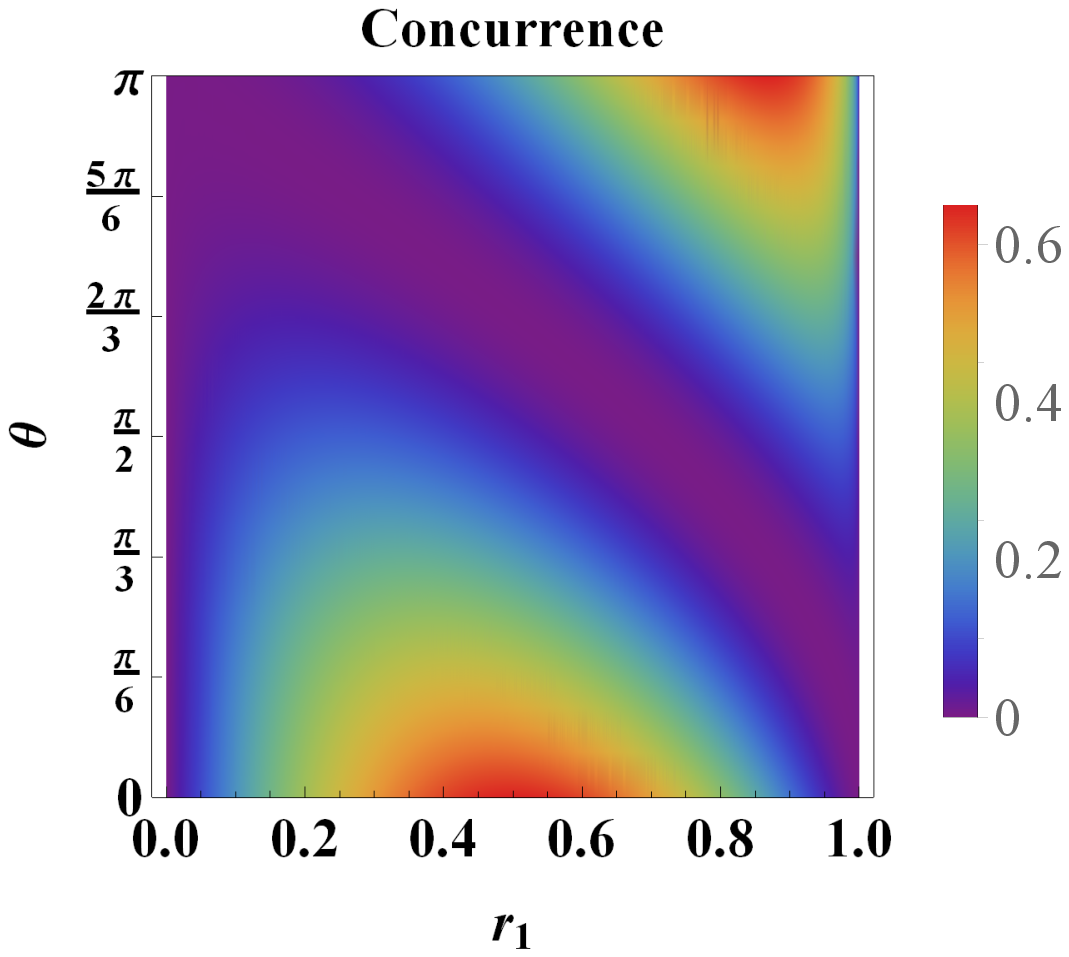}}
\hspace{0.01\textwidth}
\subfigure{\label{Fig2d}} {\includegraphics[width=0.45\textwidth]{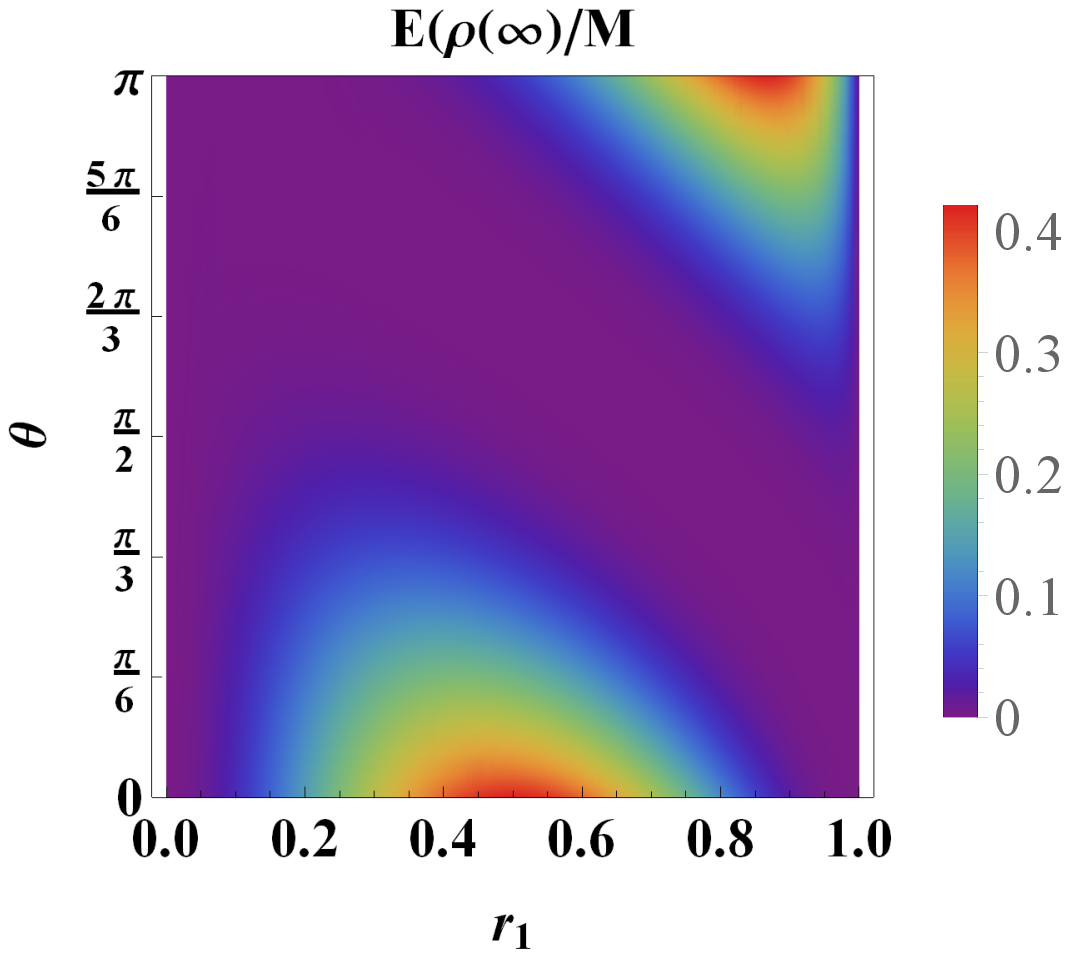}}
\hspace{0.01\textwidth}
\caption{ { (Color online) Stationary entanglement quantified by concurrence (left panels) and entanglement measure (\ref{eq:entmea}) (right panels)  versus $r_1$ and $\theta$ for $\phi=\pi$ (top panels) and  $\phi=0$ (bottom panels).} } \label{Fig2}
   \end{figure}
 In Fig. \ref{Fig2} we have compared the stationary entanglement based on Eq. (\ref{eq:entmea}) and concurrence \cite{Wootters1998}. The great agreement between these two measures allows us to use the measure defined in Eq. (\ref{eq:entmea}) in the rest of the paper. We have plotted the two measures as functions of $r_1$ and $\theta$ for two values of $\phi$. First of all, it is evident that the maximum value of entanglement is $1$ for $\phi=\pi$ whereas it is in the vicinity of $0.41$ for $\phi=0$. 

\begin{figure}[ht]
\centering
\includegraphics[width=0.5\textwidth]{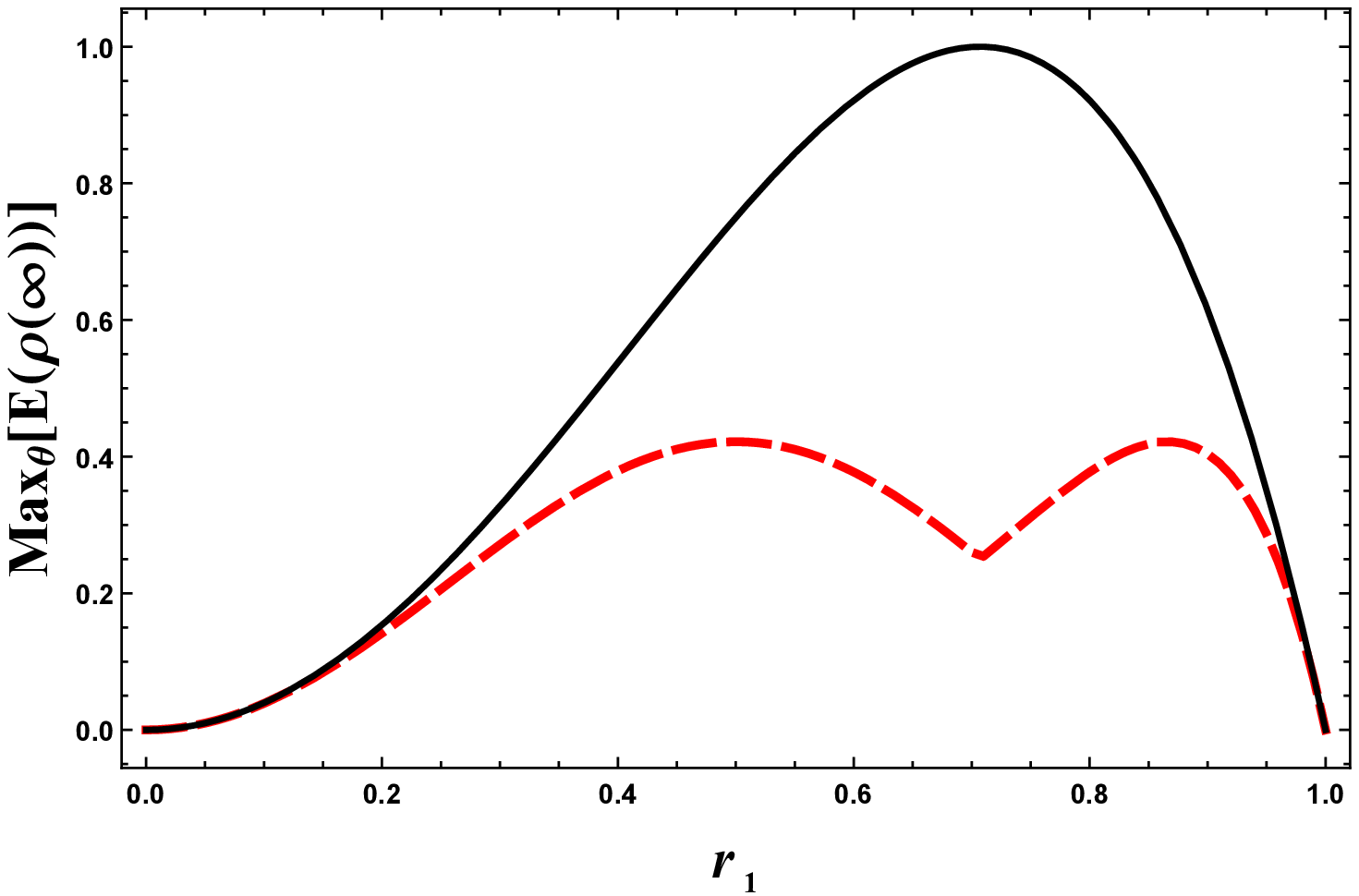}
\caption{(Color online) The maximum value of the stationary entanglement $E(\rho(\infty))/M$ as a function of $r_1$ over parameter $\theta$ for $\phi=0$ (dasshed line) and $\phi=\pi$ (solid line). We point out that the these results are completely independent of environmental variables as well as the classical driving field.} \label{Fig3}
   \end{figure}

It is evident that the region $\left\lbrace r_1,\theta\right\rbrace $ for which the stationary entanglement is near its maximum value, i.e., $E(\rho(\infty))/M=1$ is obtained for $\phi=\pi$. This is due to the fact that for $\phi=\pi$ most part of the initial state lays in the sub-radiant state with maximum degree of entanglement. For $\phi=0$ and $2\pi$, this region shrinks and the maxima are lower, as can be readily observed from Fig. \ref{Fig3}. According to Fig. \ref{Fig3}, the optimal value of the stationary entanglement depends on the relative phase $\phi$. For $\phi=0$ and/or $2\pi$, the maximum value of the stationary entanglement, i.e., $E^{\text{max}}(\rho(\infty))/M\simeq 0.41$, is obtained at $r_1\simeq 0.57$ and at $r_1\simeq 0.87$. Our further numerical calculations reveal that the value of $\theta$ for which the stationary entanglement is optimal, is $\theta=0$ at  $r_1\simeq 0.57$, while it is $\theta=\pi$ at $r_1\simeq 0.87$. We point out that both cases resemble initially separable states. It is also possible to illustrate that the optimal stationary entanglement has a local minimum at $r_1\simeq 0.71\simeq\frac{1}{\sqrt{2}}$, i.e., when the qubits are equally coupled to the environment.

For $\phi=\pi$, the maximum amount of the stationary entanglement is obtained at $r_1=r_2=\frac{1}{\sqrt{2}}$. Our numerical calculations show that the optimal value of $\theta$ (i.e., the value of $\theta$ at which the stationary entanglement is optimal) is $\theta=\frac{\pi}{2}$. This corresponds to the initial state $\ket{\psi(0)}=\frac{1}{\sqrt{2}}\left( \ket{E}\ket{G}-\ket{G}\ket{E}\right).$

\subsection*{Dynamics of Entanglement}
\label{Sec.Dyn}
Now, we are in a position to investigate the dynamics of entanglement for the case in which the two qubits have the same transition frequency. According to Eq. (\ref{eq:C}), the dynamics of the qubit system depends on the amplitude ${\cal G}(t)$. First, we note that using relation (\ref{Correlation function}), the correlation function $F(t-t')$ may be obtained as
\begin{equation}
\label{eq:kernelsol}
F(t-{{t}'})=W^2e^{-\lambda(t-{{t}'})}e^{i(\chi+\Delta_L)(t-{{t}'})} \, ,
\end{equation}
in which  $\Delta_L=\omega_L-\omega_c$ denotes the detuning between the classical driving field $\omega_L$ and central frequency of the cavity $\omega_c$.

Then in order to derive ${\cal G}(t)$, we take a Laplace transform, ($f(s)\equiv \int_{0}^{\infty} f(t) e^{-st}dt$) to both sides of Eq. (\ref{intgdiff}). This converts the integro-differential equation into an algebraic one, which can be easily solved to obtain the Laplace transformation of ${\cal G}(t)$. Then, by taking an inverse Laplace transformation, this parameter can be obtained as
\begin{equation}
{\cal G}(t)=e^{-Mt/2}\left( \cosh ({\cal F}t/2)+\frac{M}{{\cal F}}\sinh({\cal F}t/2)\right) \, ,
\end{equation}  
in which $M=\lambda -i(\chi+\Delta_L )$ and ${\cal F}=\sqrt{{{M}^{2}}-\alpha_{_T}^2W^2 {{(1+\cos \eta )}^{2}}}$.   Before considering the dynamics of entanglement, we define the dimensionless parameter $R=\frac{{\cal R}}{\lambda}$ to explore both strong and weak coupling regimes corresponding to $R\gg 1$ and $R\ll 1$, accordingly. 

{ The entanglement measure (\ref{eq:entmea}) for quantum state (\ref{eq:densitymatrix1}) takes the following analytical expression
\begin{equation}
E(\rho(t))/M=4|C_1(t)|^2|C_2(t)|^2 .
\label{eq:Constmea}
\end{equation}
}

\subsubsection*{Resonance Scenario}

\begin{figure}[ht]
\centering
\subfigure{\label{Fig4a}} {\includegraphics[width=0.4\textwidth]{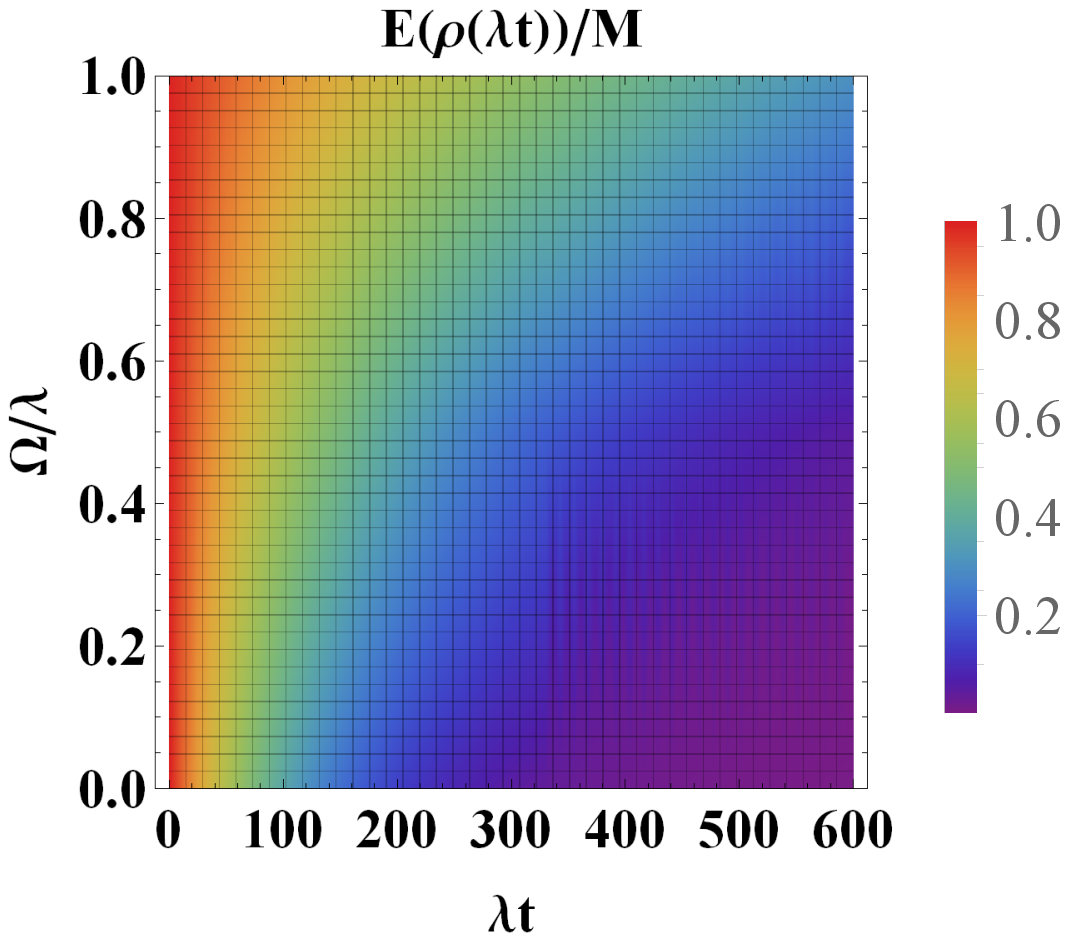}}
\hspace{0.005\textwidth}
\subfigure{\label{Fig4b}} {\includegraphics[width=0.4\textwidth]{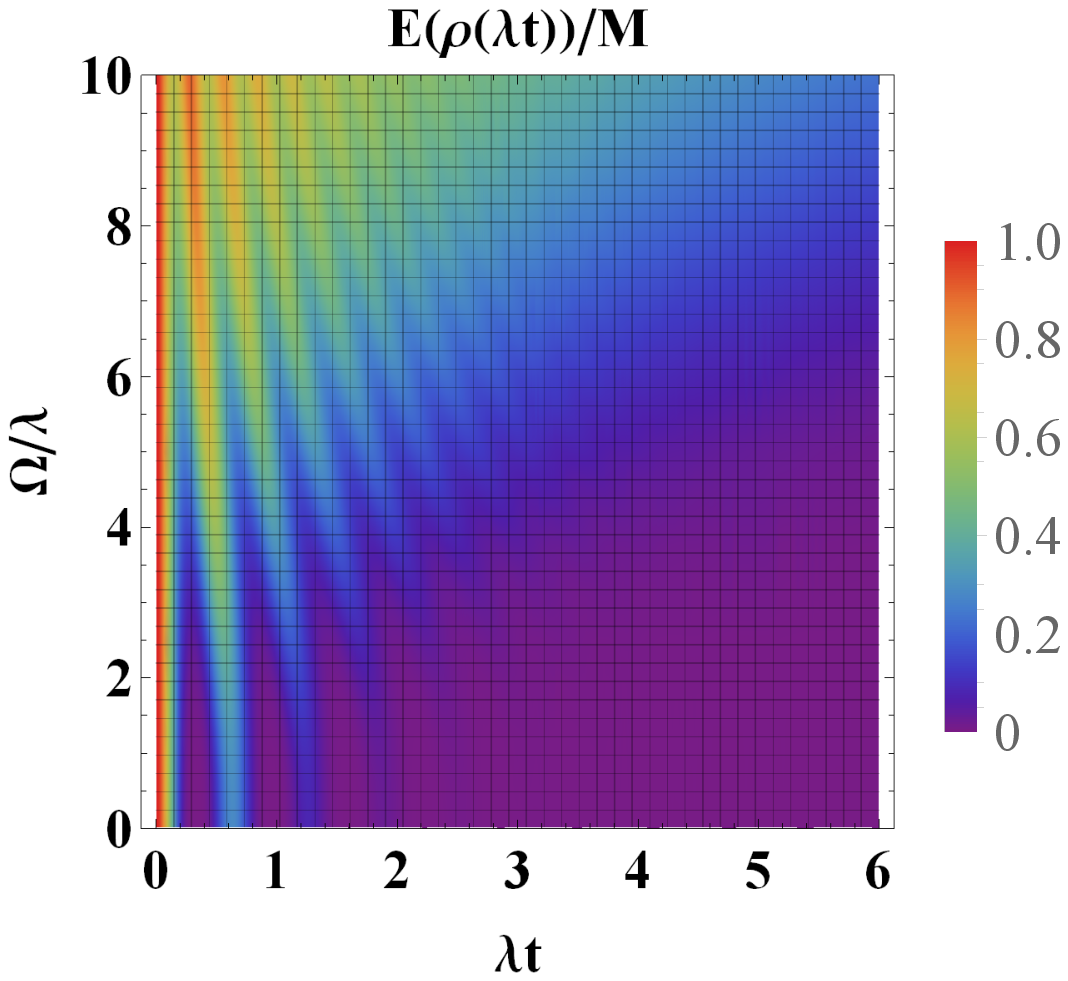}}

\caption{(Color online) Dynamics of entanglement in the presence of classical driving field for an initially entangled state for (a) weak coupling regime, i.e., $R=0.1$ and (b) strong coupling regime, i.e., $R=10$. In these plots, we have considered a symmetric coupling, i.e., $r_1=\frac{1}{\sqrt{2}}$ and we have set $\Delta=\Delta_L=0$.} \label{Fig4}
   \end{figure}

In this subsection, we consider the resonance scenario, i.e.,  $\Delta=\Delta_L=0$. Fig. \ref{Fig4} illustrates the effect of the classical driving field on the entanglement dynamics for an initially entangled state. In these plots, we have set $r_1=\frac{1}{\sqrt{2}}$. First of all, it should emphasised that in the absence of the classical driving field, the results are quite analogue of the results presented in \cite{PhysRevLett.100.090503,Nourmandipour2015}, where the measure concurrence \cite{Wootters1998} has been used to quantify the degree of entanglement. This supports our results quantified with the new measure (\ref{eq:entmea}). In this scenario, a decaying behaviour of entanglement is observed for both strong and weak coupling regimes. However, in the strong coupling regime, the oscillatory behaviour for entanglement is clearly seen due to the long memory of the environment, a phenomenon declaring the non-Markovian process. As is observed, the initial maximum value of entanglement is finally washed out at sufficiently long times for all values of the parameter $r_1$.

In the presence of classical driving field,  a different behaviour is clearly observed in both coupling regimes. For instance, in the weak coupling regime, the classical driving makes the decay of entanglement slower. However, the entanglement vanishes at sufficiently long times. In the strong coupling regime, the entanglement sudden death is clearly gone due to the influence of the classical driving field. According to the presented plots, the positive role of the classical driving field on the preservation of entanglement is clearly observed: the larger the Rabi frequency, the better preservation of the initial entanglement.  

\begin{figure}[ht]
\centering
\subfigure{\label{Fig5a}} {\includegraphics[width=0.4\textwidth]{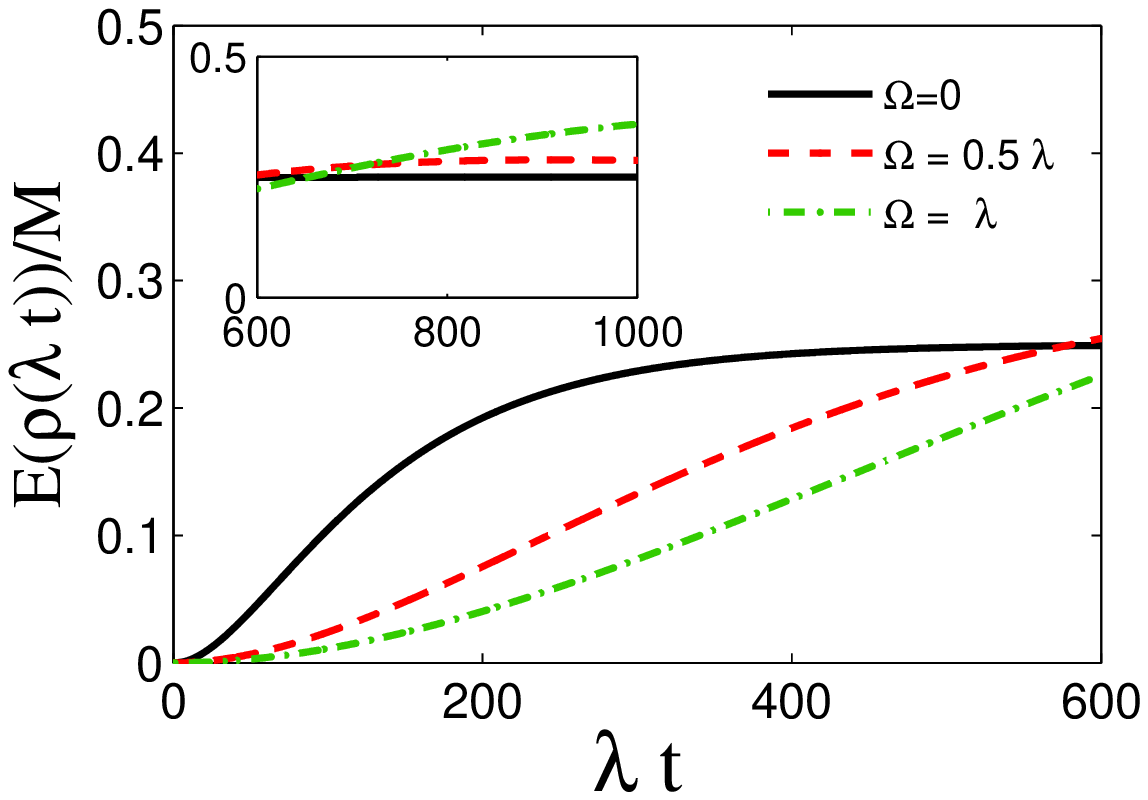}}
\hspace{0.005\textwidth}
\subfigure{\label{Fig5b}} {\includegraphics[width=0.4\textwidth]{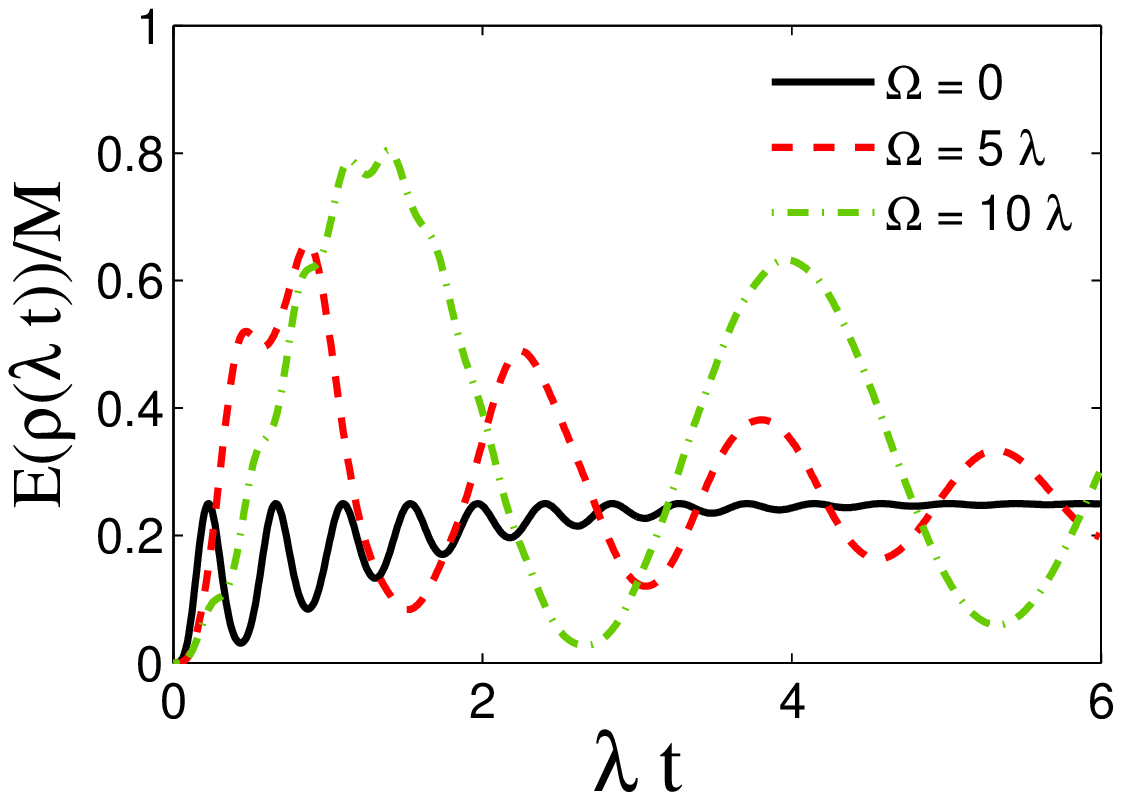}}

\caption{(Color online) Dynamics of entanglement in the presence of classical driving field for an initially separable state for (a) weak coupling regime, i.e., $R=0.1$ and (b) strong coupling regime, i.e., $R=10$. In these plots, we have considered a symmetric coupling, i.e., $r_1=\frac{1}{\sqrt{2}}$ and we have set $\Delta=\Delta_L=0$.} \label{Fig5}
   \end{figure}

So far, we have explored the role of the classical driving field on the preservation of the initial entanglement between the two qubits. It is also interesting to examine the influence of the classical driving field on the entanglement when the two qubits are initially separable. Fig. \ref{Fig5} illustrates the entanglement dynamics for an initially separable state under the influence of the classical driving field. It is interesting to notice the effect of the classical driving field in the strong coupling regime. In the absence of the classical driving field, the entanglement never exceeds its stationary value. However, the classical driving field makes the entanglement even greater than its stationary value with a maximum of around $0.8$ for $\Omega=10\lambda$. The inset in Fig. \ref{Fig5a} shows that it is also possible to exceed the stationary value of entanglement for the weak coupling regime. Again, from these plots, it is evident that the stationary entanglement is completely independent of environmental variables as well as the classical driving field which is completely in agreement with Fig. \ref{Fig3}.

\subsubsection*{Off-Resonance Scenario}

In this subsection we discuss the off-resonance scenario. We begin by considering the case in which the detuning between the classical driving field and the central frequency of the cavity is zero, i.e., $\Delta_L=0$.

As is observed from Fig. \ref{Fig4}, the initial entanglement is washed out at sufficiently long (scaled) times. Therefore, to investigate the role of detuning parameter, we have plotted the entanglement measure versus $\Omega$ and the detuning parameter $\Delta$ at a certain value for scaled time, i.e., $\lambda t=400$ for weak coupling regime and $\lambda t=4$ for strong coupling regime, at which, the initial entanglement has already been disappeared (in the absence of classical driving field as well as the detuning parameter). First of all, it is evident that for small values of detuning, the behaviour of entanglement is similar to the case of the resonance. However, for larger values of detuning, the entanglement shows a monotonic increase toward its maximum value. According to the effective Hamiltonian (\ref{Heff}),  increasing the value of $\Delta$  enhances the effective transition frequency (energy) of the qubits under the driving field. This makes qubits to be robust against the influence of the environment.

\begin{figure}[ht]
\centering
\subfigure{\label{Fig6a}} {\includegraphics[width=0.4\textwidth]{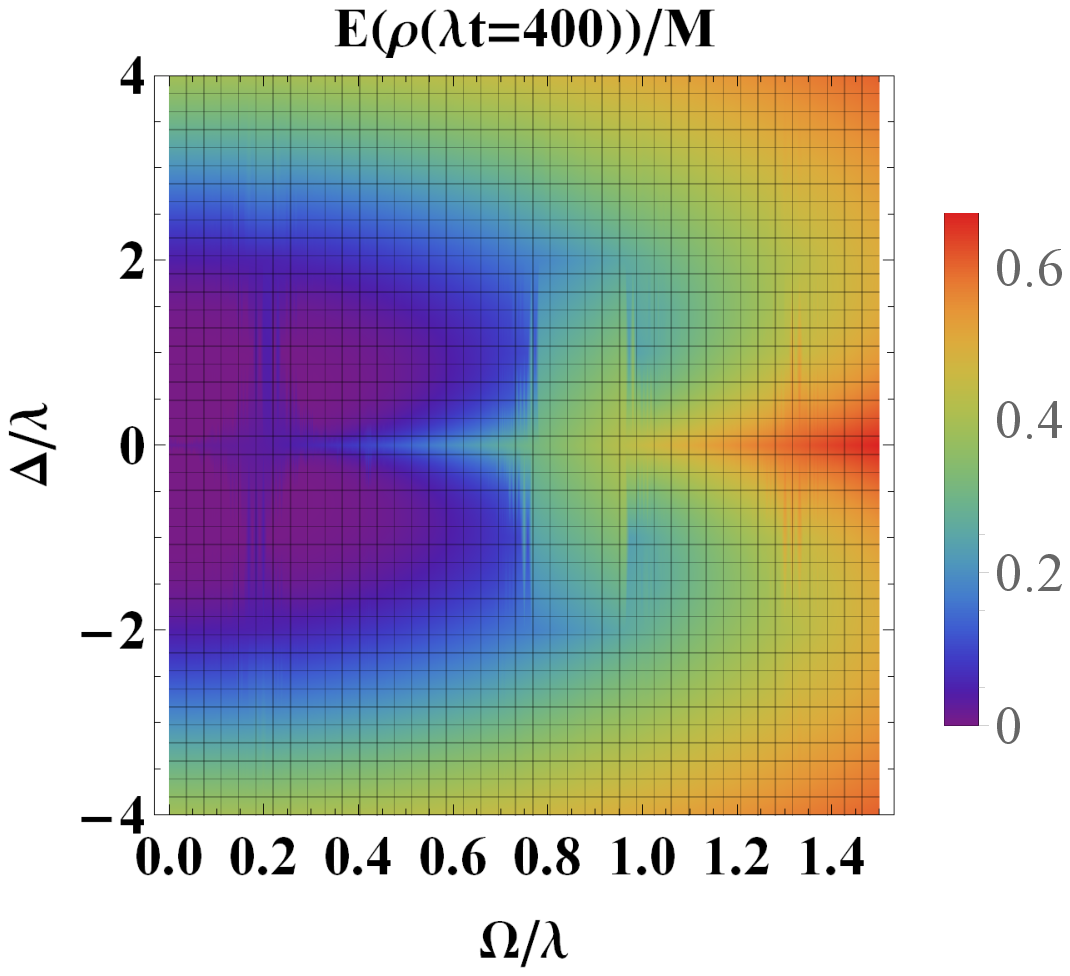}}
\hspace{0.005\textwidth}
\subfigure{\label{Fig6b}} {\includegraphics[width=0.4\textwidth]{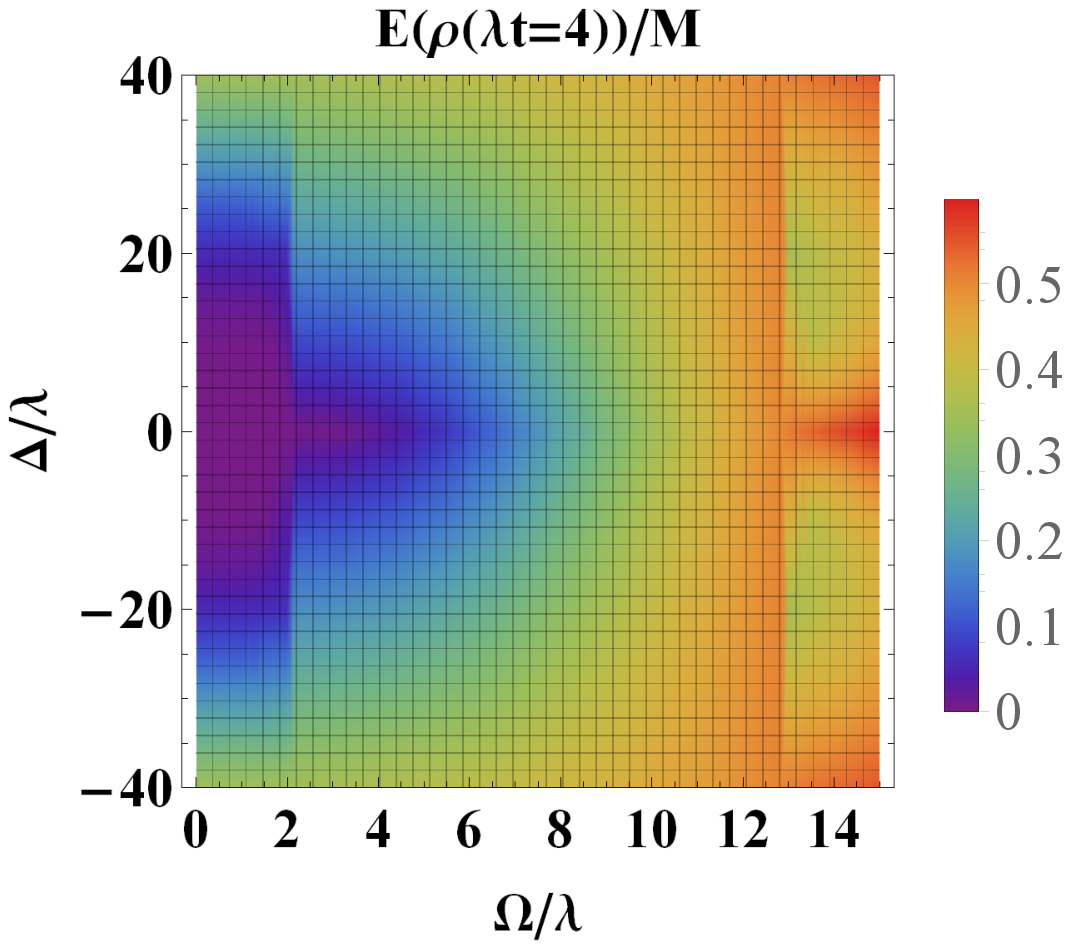}}

\caption{(Color online) Entanglement versus the driving field $\Omega$ and the detuning parameter $\Delta$ for an initially entangled state  at scaled time (a) $\lambda t=400$ for weak coupling regime, i.e., $R=0.1$ and (b) $\lambda t=4$ for strong coupling regime, i.e., $R=10$. Here we have set $\Delta_L=0$.} \label{Fig6}
   \end{figure}

\begin{figure}[ht]
\centering
\subfigure{\label{Fig7a}} {\includegraphics[width=0.4\textwidth]{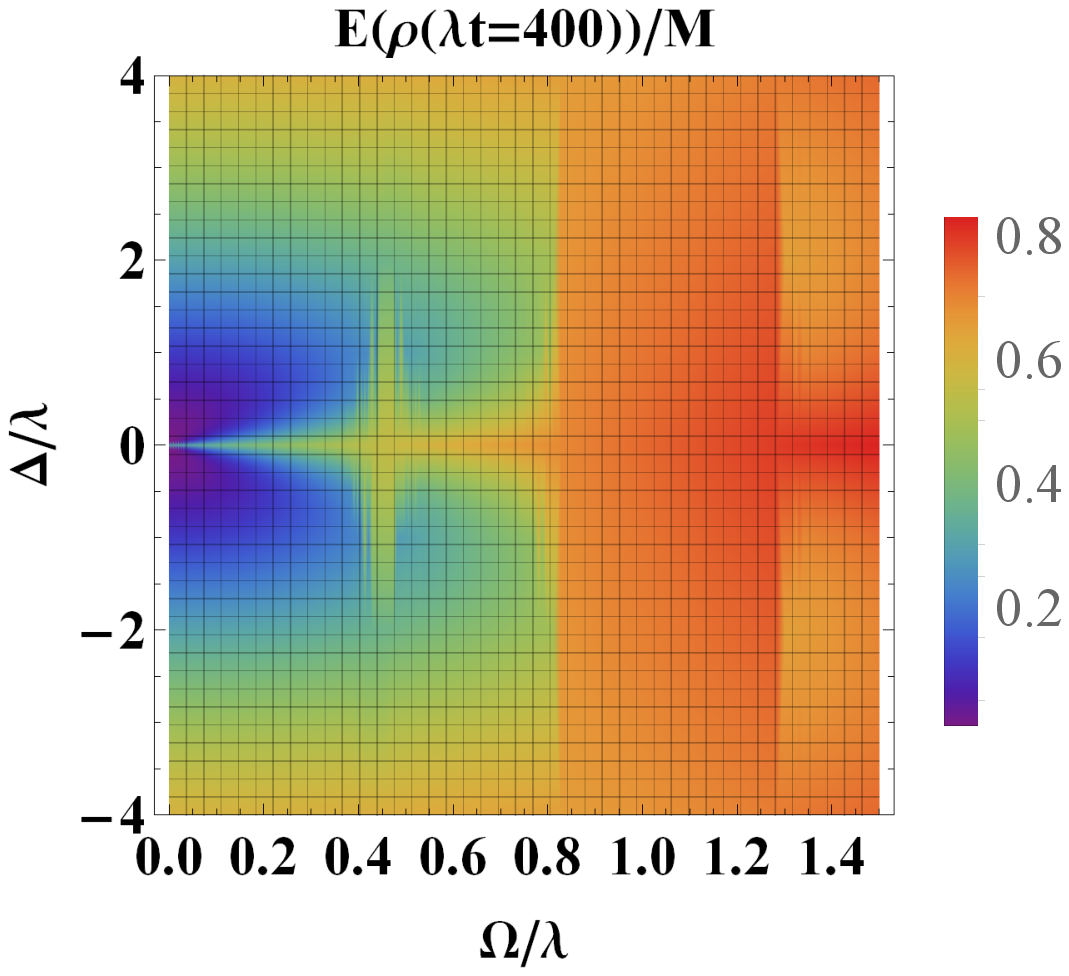}}
\hspace{0.005\textwidth}
\subfigure{\label{Fig7b}} {\includegraphics[width=0.4\textwidth]{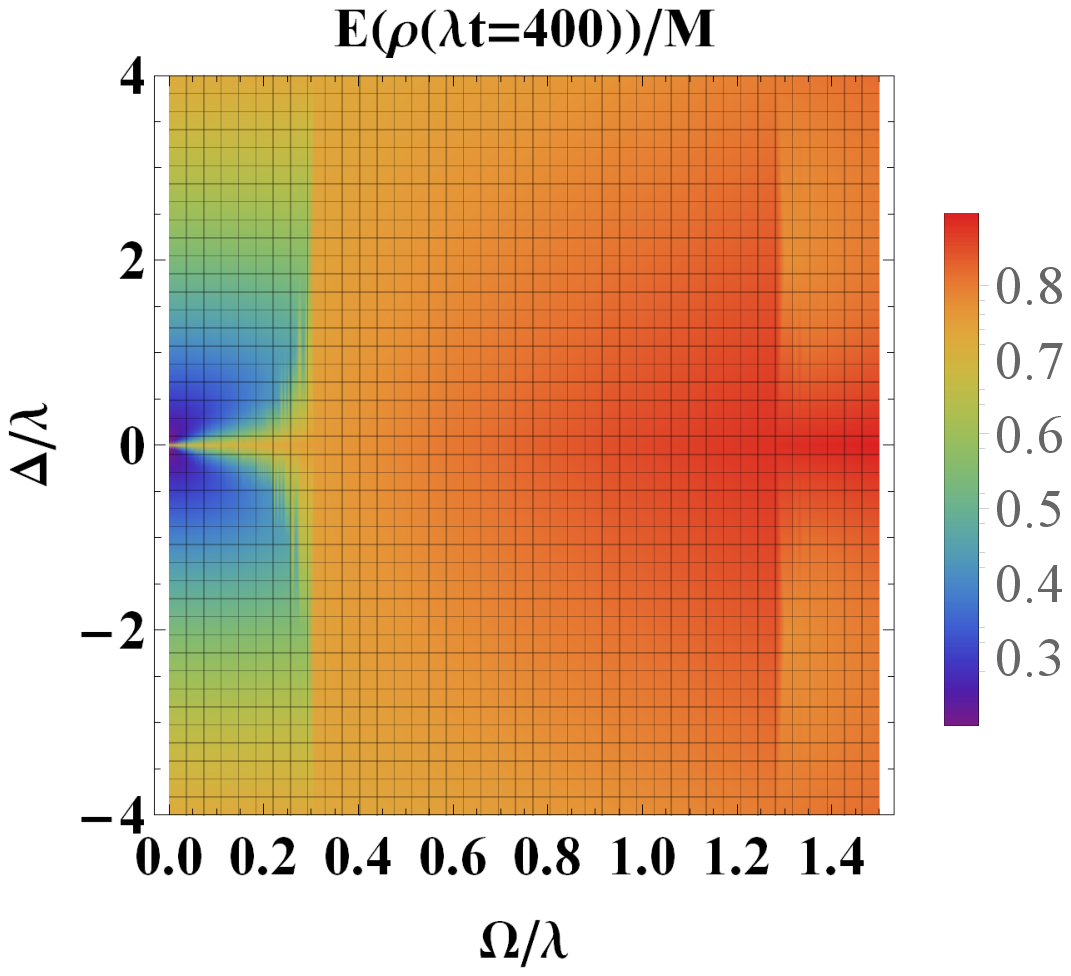}}
\caption{(Color online) Entanglement versus the driving field $\Omega$ and the detuning parameter $\Delta$ for an initially entangled state in the weak coupling regime, i.e., $R=0.1$ at scaled time $\lambda t=400$ with (a) $\Delta_L=1.5\lambda$ and (b) $\Delta_L=3\lambda$.} \label{Fig7}
   \end{figure}

In Fig. \ref{Fig7} we have assumed non-zero values for the detuning between the classical driving field and the central frequency of the cavity (i.e., $\Delta_L\neq 0$) when the system is in the weak coupling regime. Other parameters are similar to Fig. \ref{Fig6}. It is evident the constructive role of detuning parameter $\Delta_L$ on the preservation of the initial entanglement. Specially, for large values of $\Delta_L$, even for small values of $\Delta$  and driving filed, the entanglement survives much better. Our other numerical calculations illustrate that the same behaviour is observed for large values of $\Delta_L$. The same behaviour can be observed for strong coupling regime. 

\subsection*{Improvement of the Preservation of Entanglement}

As is observed in the previous subsection, the initial amount of entanglement between two qubits tends to zero at sufficient long times. Although the classical driving field has a constructive role in preserving the initial entanglement, it does not completely preserve the initial entanglement. Here we intend to examine the possible role of an interaction term among the two qubits on the entanglement dynamics \cite{Rafiee2017}. To this end, we consider the following extra Hamiltonian in relation (\ref{Heff})
\begin{equation}
\hat{H}_{\text{int}}=2J\hat{\Varrho}_{z}^{(1)}\hat{\Varrho}_{z}^{(2)}
\end{equation}
in which $J$ is the coupling constant. 

We again assume an initial state of the form (\ref{eq:initialstate}). Consequently, the quantum state of the entire system+environment is 
\begin{equation}
\ket{\psi(t)}=C_{1}^J(t)e^{2iJt}\ket{E}\ket{G}\ket{\boldsymbol{0}}_{R}+C_{2}^J(t)e^{2iJt}\ket{G}\ket{E}\ket{\boldsymbol{0}}_{R}+\sum_{k}C_{k}^J(t)e^{-2iJt}e^{-i(\omega_k-\chi)t}\ket{G}\ket{G}\ket{\boldsymbol{1}_k} \, .
\label{eq:stateJ}
\end{equation}
The procedure for solving the Schr\"{o}dinger equation is exactly the same approach presented before. It is quite straightforward to show that the kernel function (\ref{eq:kernelsol}) now takes the form $
F_J(t-{{t}'})=W^2e^{-\lambda(t-{{t}'})}e^{i(\chi+\Delta_L-4J)(t-{{t}'})} \, .$ In this case, the new survival amplitude of the super-radiant state is obtained as
\begin{equation}
{\cal G}_J(t)=e^{-M_Jt/2}\left( \cosh ({\cal F_J}t/2)+\frac{M_J}{{\cal F_J}}\sinh({\cal F_J}t/2)\right) \, ,
\end{equation}  
in which $M_J=\lambda -i(\chi+\Delta_L -4J)$ and ${\cal F}_J=\sqrt{{{M_J}^{2}}-\alpha_{_T}^2W^2 {{(1+\cos \eta )}^{2}}}$. Finally, the amplitudes $C_i^J(t)$ result
\begin{equation}
\label{eq:CJ}
C_{1}^J(t)=r_2\beta_-+r_1{\cal G}_J(t)\beta_+ \ \ \ \text{and} \ \ \  C_{2}^J(t)=-r_1\beta_-+r_2{\cal G}_J(t)\beta_+
\end{equation}
in which $\beta_-$ and $\beta_+$ have been defined before. 

\begin{figure}[ht]
\centering
\subfigure{\label{Fig8a}} {\includegraphics[width=0.4\textwidth]{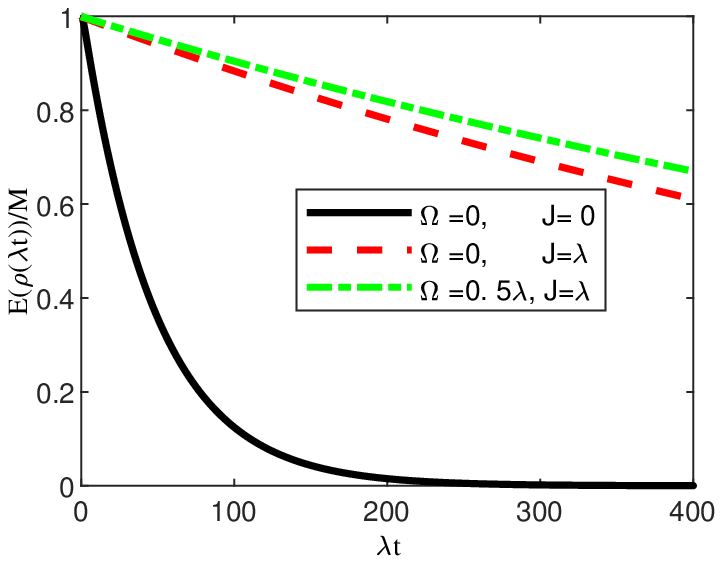}}
\hspace{0.005\textwidth}
\subfigure{\label{Fig8b}} {\includegraphics[width=0.4\textwidth]{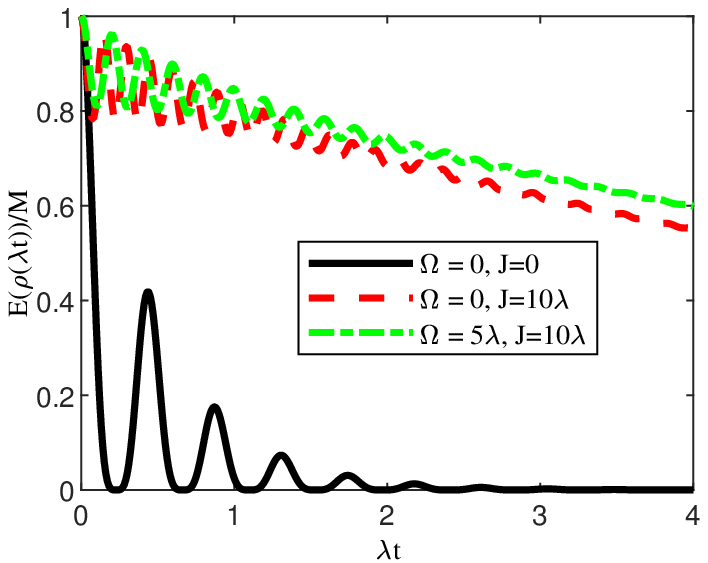}}
\caption{(Color online) 
Dynamics of entanglement in the presence of classical driving field and an interaction term for an initially entangled state for (a) weak coupling regime, i.e., $R=0.1$ and (b) strong coupling regime, i.e., $R=10$. In these plots, we have considered a symmetric coupling, i.e., $r_1=\frac{1}{\sqrt{2}}$ and we have set $\Delta=\Delta_L=0$.} \label{Fig8}  
   \end{figure}

Figure \ref{Fig8} illustrates the entanglement in the presence of both classical driving filed and the interaction Hamiltonian between the two qubits. It is evident the positive effect of the parameter $J$ on the entanglement preservation. According to these plots, its protecting role is even more efficient that the classical driving field. 

\section*{Dissimilar qubits} 

Now we pay attention to a different scenario in which the two qubits do not have similar transition frequencies, i.e., $\omega_1\neq\omega_2$. In this scenario, no subradiant (decoherence-free) state exists. In order to find the analytical solution, we first take the Laplace transformation from both sides of equations (\ref{eq:newdotu}) to arrive at the following algebraic equations for $\tilde{C}_1(s)$ and $\tilde{C}_2(s)$:
\begin{subequations}
\label{eq:lapdotu}
\begin{eqnarray}
s\tilde{C}_1(s)-C_1(0)&=&-\left[ \alpha_1^2{{\cos }^{4}}(\eta_1 /2)\tilde{C}_1(s) + \alpha_1\alpha_2{{\cos }^{2}}(\eta_1 /2){{\cos }^{2}}(\eta_2 /2)\tilde{C}_2(s+i\xi_{21})\right]\tilde{G}(s-i\xi_1) \, , \label{eq:lapu1} \\
s\tilde{C}_2(s)-C_2(0)&=&-\left[\alpha_1\alpha_2{{\cos }^{2}}(\eta_1 /2){{\cos }^{2}}(\eta_2 /2)\tilde{C}_1(s-i\xi_{21}) +\alpha_2^2{{\cos }^{4}}(\eta_2 /2)\tilde{C}_2(s)\right]\tilde{G}(s-i\xi_2) \, . \label{eq:lapu2} 
\end{eqnarray}
\end{subequations}
Then, by solving the above set of equations we get
\begin{equation}
\label{CLap}
\tilde{C}_j(s)=\tilde{{\cal G}}_{j1}(s,r_1)C_1(0)+\tilde{{\cal G}}_{j2}(s,r_1)C_2(0) \, , \ \ \ \ (j=1,2)
\end{equation}
in which,
\begin{equation}
\begin{aligned}
\label{lapG}
\tilde{{\cal G}}_{jj}(s,r_1)&=\frac{s^2+a_js+b_j}{s^3+A_js^2+B_js+C_j}  \, ,  \ \ \ \ \ \text{and}  \\
\tilde{{\cal G}}_{ji}(s,r_1)&=\frac{c_j}{s^3+A_js^2+B_js+C_j} \, , \ \ \ \ \ j\neq i \, .
\end{aligned}
\end{equation}
Here
\begin{equation}
\begin{aligned}
a_{1,2}&=\lambda-i(2\chi_{1,2}-\chi_{2,1}+\Delta_L) \, , \\
b_{1,2}&=(\chi_{2,1}-\chi_{1,2})(i\lambda+\chi_{1,2}+\Delta_L)+{\cal R}^2r_{2,1}^2\cos^4(\eta_{2,1}/2)
\, , \\
c_{1,2}&=-{\cal R}^2r_{1}r_{2}\cos^2(\eta_{1}/2)\cos^2(\eta_{2}/2) \, , \\
A_{1,2}&=\lambda-i(2\chi_{1,2}-\chi_{2,1}+\Delta_L) \, , \\
B_{1,2}&=(\chi_{2,1}-\chi_{1,2})(i\lambda+\chi_{1,2}+\Delta_L)+{\cal R}^2(r_{1}^2\cos^4(\eta_{1}/2)+r_{2}^2\cos^4(\eta_{2}/2)) \, , \\
C_{1,2}&=-i{\cal R}^2r_{1,2}^2(\chi_{1,2}-\chi_{2,1})\cos^4(\eta_{1,2}/2) \, .
\end{aligned}
\end{equation}
Then, we take the inverse Laplace transformation of (\ref{CLap}) to obtain the following analytical solution for the amplitude coefficients
\begin{equation}
\begin{aligned}
C_1(t)&={\cal G}_{11}(t,r_1)C_1(0)+{\cal G}_{12}(t,r_1)C_2(0) \, ,\\
C_2(t)&={\cal G}_{21}(t,r_1)C_1(0)+{\cal G}_{22}(t,r_1)C_2(0) \, ,
\end{aligned}
\end{equation}
in which, the functions ${\cal G}_{ij}(t,r_1)$ are inverse Laplace transformed of (\ref{lapG}) given as
\begin{equation}
{\cal G}_{jj}(t,r_1)=\frac{s_{j1}^2+a_js_{j1}+b_j}{(s_{j1}-s_{j2})(s_{j1}-s_{j3})}e^{s_{j1}t}-\frac{s_{j2}^2+a_js_{j2}+b_j}{(s_{j1}-s_{j2})(s_{j2}-s_{j3})}e^{s_{j2}t}+\frac{s_{j3}^2+a_js_{j3}+b_j}{(s_{j1}-s_{j3})(s_{j2}-s_{j3})}e^{s_{j3}t} \, ,
\end{equation}
and 
\begin{equation}
{\cal G}_{ji}(t,r_1)=\frac{c_j}{(s_{j1}-s_{j2})(s_{j1}-s_{j3})(s_{j2}-s_{j3})}\left[(s_{j2}-s_{j3})e^{s_{j1}t}-(s_{j1}-s_{j3})e^{s_{j2}t}+(s_{j1}-s_{j2})e^{s_{j3}t}\right] \, .
\end{equation}
In the above relations, $s_{j1}$, $s_{j2}$ and $s_{j3}$ are the roots of the following cubic equation
\begin{equation}
\label{cubic}
s^3+A_js^2+B_js+C_j=0 \ \ \ \ \ (j=1,2).
\end{equation}
We first notice that since the general cubic equation (\ref{cubic}) can be solved analytically, it is always
possible to obtain the analytical expressions of $s_{j1}$, $s_{j2}$ and $s_{j3}$ and consequently find the analytical expression for amplitude coefficients. However, these expressions are too long to be presented here. Furthermore, for the special case $\omega_1=\omega_2$ (corresponding to the sub-radiant scenario), it is straightforward to illustrate that this rather complicated analytical solution reduces to the simple expression (\ref{eq:C}). 
\begin{figure}[ht]
\centering
\subfigure{\label{Fig9a}} {\includegraphics[width=0.4\textwidth]{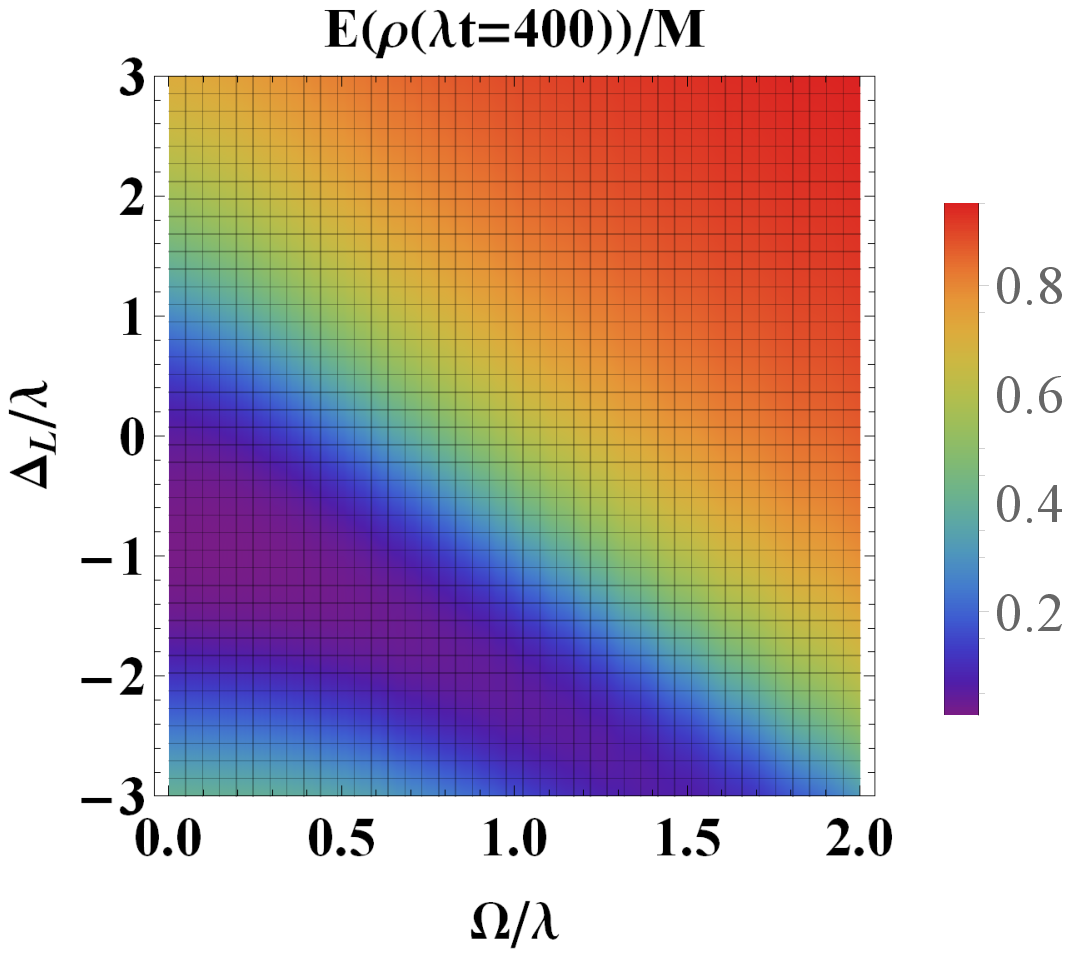}}
\hspace{0.005\textwidth}
\subfigure{\label{Fig9b}} {\includegraphics[width=0.4\textwidth]{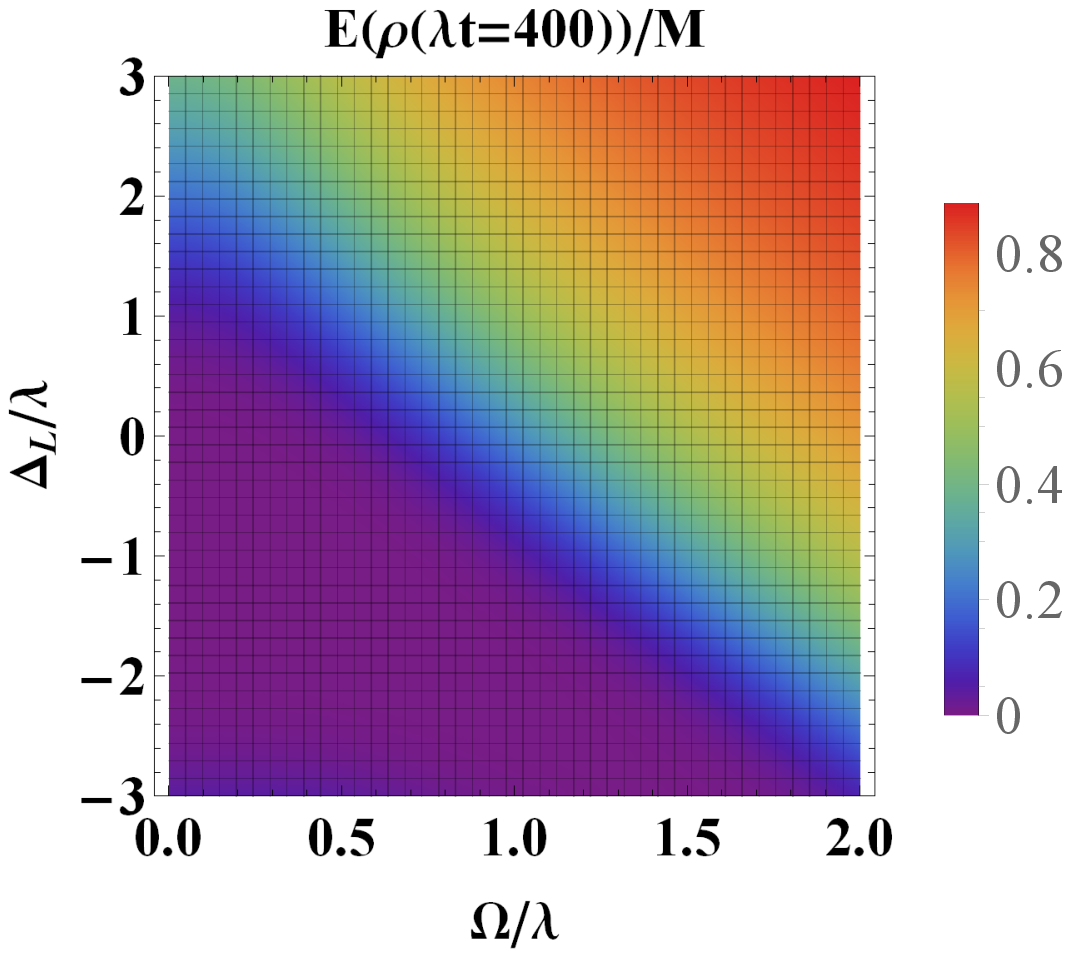}}

\caption{(Color online) Entanglement versus the driving field $\Omega$ and the detuning parameter $\Delta_L$ for an initially entangled state in the weak coupling regime, i.e., $R=0.1$ at scaled time $\tau=400$ with (a) $\Delta_1=0$, $\Delta_2=\lambda$ and (b) $\Delta_2=-\Delta_1=\lambda$.} \label{Fig9}
   \end{figure}

\begin{figure}[ht]
\centering
\subfigure{\label{Fig10a}}{\includegraphics[width=0.4\textwidth]{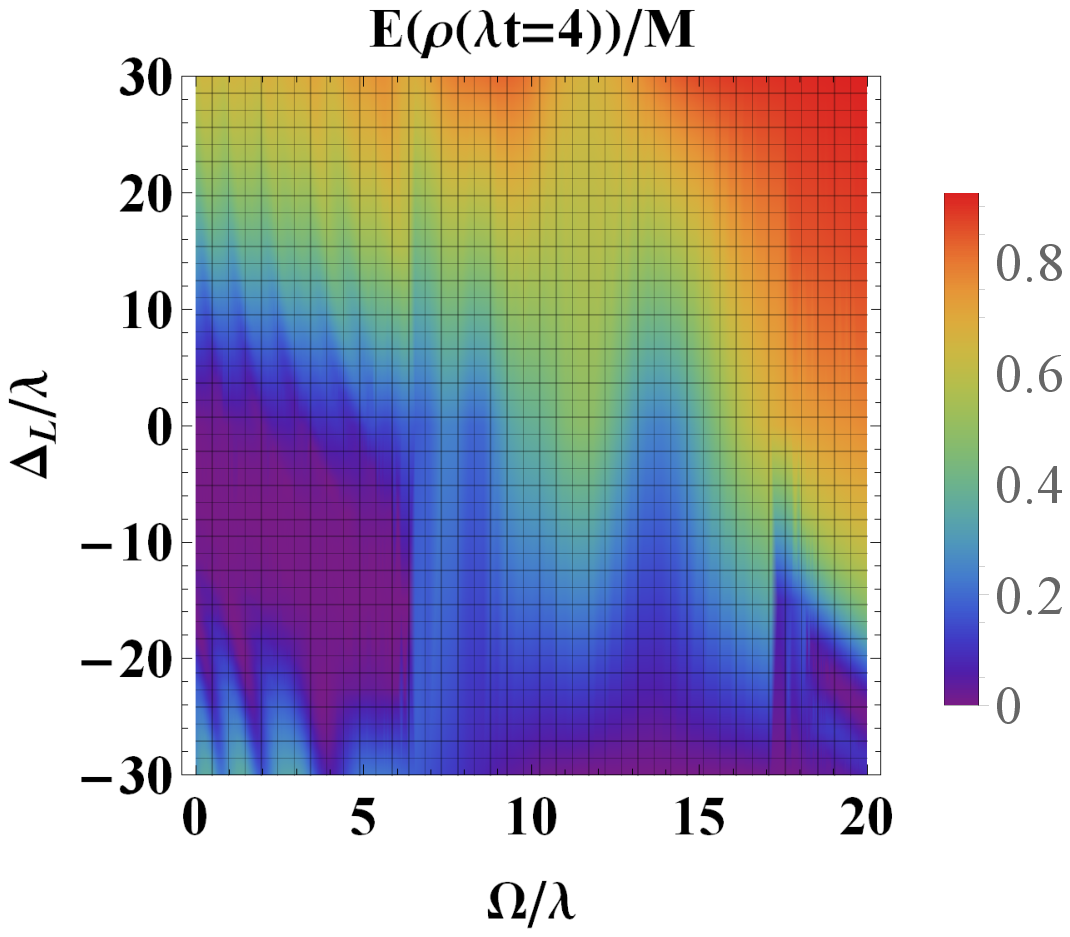}}
\hspace{0.005\textwidth}
\subfigure{\label{Fig10b}} {\includegraphics[width=0.4\textwidth]{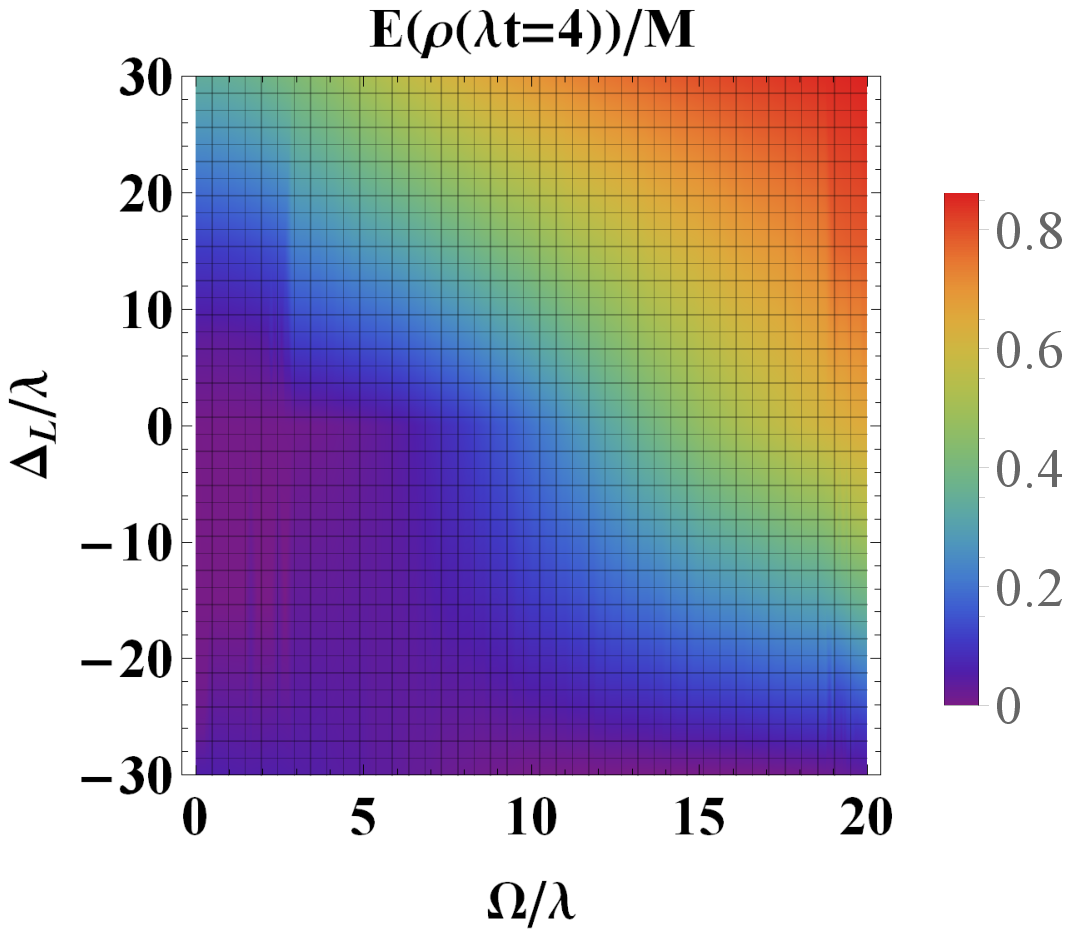}}

\caption{(Color online) Entanglement versus the driving field $\Omega$ and the detuning parameter $\Delta_L$ for an initially entangled state in the strong coupling regime, i.e., $R=10$ at scaled time $\tau=4$ with (a) $\Delta_1=0$, $\Delta_2=10\lambda$ and (b) $\Delta_2=-\Delta_1=10\lambda$.} \label{Fig10}
   \end{figure}

In Fig. \ref{Fig9} we have plotted the entanglement in the weak coupling regime at scaled time $\tau=400$ versus the detuning $\Delta_L$ achieved starting from a maximally entangled initial-state and for various values of detuning parameters. According to the information supplied, $\Delta_L$ must take positive values to have a constructive role in the preservation of entanglement. This means that the frequency of the classical driving field must be greater than the center frequency of the cavity. However, for the case in which $\Delta_1$ is close to $\Delta_2$, for large negative values of $\Delta_L$, we observe a positive role in the preservation of entanglement. The same behavior is seen in the case of the strong coupling regime (see Fig. \ref{Fig10}).

\section*{Conclusions}

To sum up, we have investigated the analytical dynamics of entanglement for two qubits dissipating into a common environment. The two qubits can have different transition frequencies and are driven by a classical driving field. We have investigated the dynamics of entanglement without restricting ourselves to the Born-Markov approximation. We have used the new entanglement measure defined in \cite{PhysRevA.101.042129} rather than concurrence which is usually used for two-qubit systems. The availability of the new entanglement measure enables us to investigate the entanglement dynamics for an arbitrary number of qubits \cite{Nourmandipour2016} in a common environment. 

We first considered the case in which the qubits have the same transition frequency. In this case, we illustrated in detail that there exists a sub-space that does not evolve in time. Then, it is possible to find the stationary state of entanglement at sufficiently long times. The surprising aspect here is that the stationary state does not depend on the environment properties as well as the classical driving field. Base on this fact, we investigated the stationary state of entanglement as a function of the coupling rate of $r_1$ and the initial state. Particularly, we determined the situation in which a maximally entangled state can be obtained even after interaction with the environment. 

Then we investigated the impact of the classical driving field on the dynamics of entanglement for two similar qubits. In the absence of any detuning and for an initially entangled state, the entanglement has a decaying behaviour as time goes on. However, oscillatory behaviour is seen in the strong coupling regime due to the memory depth of the environment. The constructive role of the classical driving field is clearly seen in Figs. \ref{Fig4}. Especially, in the strong coupling regime and for large values of $\Omega$, the oscillatory behaviour disappears. On the other hand, we illustrated that an initially factorable state can be entangled and even persists at a steady state via interaction with the common environment. As is stated before, the stationary entanglement does not depend on the classical driving field. However, our results illustrate that this stationary value of entanglement can be exceeded in the presence of the classical driving field. Particularly, in the strong coupling regime and for sufficiently large values of the amplitude of the classical driving field, a high degree of entanglement (i.e., $E(\rho(\infty))/M=0.8$) can be achieved from an initially disentangled state (see Fig. \ref{Fig5}b). Furthermore, we have investigated the role of detuning on the survival of initial entanglement in both weak and strong coupling regimes. Overall, the detuning has a constructive role in the preservation of entanglement. First, in the absence of detuning between the frequency of the classical driving field and the central frequency of the cavity, i.e., ($\Delta_L=0$), the detuning parameter $\Delta$ can play a constructive role. More surprisingly, for nonzero values of $\Delta_L$, we observe a much better performance of preservation. We also illustrated that an interaction term between the two qubits can boost the preservation process.

We should state that the new technologies in the quantum era compel us to miniature the physical devices as much as possible, regardless of the presence or absence of direct subsystem interactions \cite{PhysRevB.77.155420}. This requires generating, control, and even enhance the entanglement in those systems in which the environmental effects can not be ignored. Therefore, we expect the presented
analytical results would be the first step towards that goal.  Especially, the introduced controlling method (utilizing the classical driving field) can be easily implemented in nowadays experiments. Whereas, there are difficulties in using other entanglement protection methods from the practical point of view. For instance, in the quantum Zeno effect, it is hard to perform the non-selective measurements to freeze the state of the system in a frozen subspace. Furthermore, in some situations, these non-selective measurements not only do not protect the entanglement but also speed up its decay, a phenomenon called the anti-quantum Zeno effect \cite{Nourmandipour2016J}.

Furthermore, this model can be generated to an array of qubits with an arbitrary number of qubits driven by a classical field in a lossy cavity. Therefore, the new multi-partite entanglement measure paves the way for quantifying multi-partite entanglement analytically. This allows us to investigate the quantum synchronization between two clusters of qubits not only in the Markovian regime \cite{PhysRevLett.113.154101} but also in the non-Markovian regime. On the other hand, by considering an interaction among the qubits (for instance an Ising model), the new entanglement measure can be used to investigate dynamical quantum phase transition. This will shed light on the relation between quantum phase transition and quantum entanglement in multi-partite systems. Also, we should point out that in a multi-partite system, by ignoring the condition $\Omega <<{{\omega }_{j}},{{\omega }_{L}}$, the introduced unitary transformation leads to nonlinear terms which enable one to generate spin squeezing in the presence of dissipation. Then, the introduced measure determines the relation between entanglement and spin squeezing. These are left for future works. Finally, from a practical point of view, the trapped ions coupled to the bath of vacuum modes of the radiation field could be a suitable candidate as an experimental implementation \cite{Barreiro2011open}.

\bibliography{mybib}

\section*{Acknowledgements}

A.N., A.V. and R.F. acknowledge support by the QuantERA ERA-NET Co-fund 731473 
(Project Q-CLOCKS). R. F. acknowledges support by National Group of Mathematical Physics (GNFM-INdAM).

\section*{Author contributions statement}

The authors have equally contributed to the manuscript. They have all read and approved its final version.

 \section*{Conflict of interest}
 The authors declare that they have no conflict of interest.
 
\section*{Additional information}
\textbf{Correspondence} and requests for materials should be addressed to A.N.

\end{document}